\documentclass[12pt, preprint]{aastex}

\shorttitle{\emph{FUSE} Observations of SwSt~1}
\shortauthors{Sterling et~al.}

\begin{document}

\title{The \emph{FUSE}\footnote{Based on observations made with the NASA-CNES-CSA \emph{Far Ultraviolet Spectroscopic Explorer}.  \emph{FUSE} is operated for NASA by the Johns Hopkins University under NASA contract NAS5-32985.}~~Spectrum of the Planetary Nebula SwSt~1: Evidence for Inhomogeneities in the Gas and Dust}

\author{N. C. Sterling\altaffilmark{2}, Harriet L. Dinerstein\altaffilmark{2}, C. W. Bowers\altaffilmark{3}, and Seth Redfield\altaffilmark{2, 4}}

\altaffiltext{2}{The University of Texas, Department of Astronomy, 1 University Station, C1400,
    Austin, TX 78712-0259; sterling@astro.as.utexas.edu, harriet@astro.as.utexas.edu, 
    sredfield@astro.as.utexas.edu}
\altaffiltext{3}{Laboratory for Astronomy and Solar Physics, Code 681,
   NASA Goddard Space Flight Center, Greenbelt, MD 20771; bowers@stis.gsfc.nasa.gov}
\altaffiltext{4}{Harlan J.\ Smith Postdoctoral Fellow at McDonald Observatory}

\begin{abstract}

We present \emph{Far Ultraviolet Spectroscopic Explorer} (\emph{FUSE}) observations of the young, compact planetary nebula (PN) SwSt~1 along the line of sight to its central star HD~167362.  We detect circumstellar absorption lines from several species against the continuum of the central star.  The physical parameters of the nebula derived from the \emph{FUSE} data differ significantly from those found from emission lines.  We derive an electron density $n_{\rm e}=8800^{+4800}_{-2400}$~cm$^{-3}$ from the column density ratio of the excited \ion{S}{3} fine structure levels, which is at least a factor of 3 lower than all prior estimates.  The gaseous iron abundance derived from the UV lines is quite high ([Fe/S]~=~$-0.35\pm0.12$), which implies that iron is not significantly depleted into dust.  In contrast, optical and near-infrared emission lines indicate that Fe is more strongly depleted: [Fe/H]$=-1.64\pm0.24$ and [Fe/S]$=-1.15\pm0.33$.  We do not detect nebular H$_2$ absorption, to a limit $N$(H$_2)<7\times10^{14}$~cm$^{-2}$, at least four orders of magnitude lower than the column density estimated from infrared H$_2$ emission lines.  Taken together, the lack of H$_2$ absorption, low $n_{\rm e}$, and high gaseous Fe abundance derived from the \emph{FUSE} spectrum provide strong evidence that dense structures (which can shield molecules and dust from the destructive effects of energetic stellar photons) are not present along the line of sight to the central star.  On the other hand, there is substantial evidence for dust, molecular material, and dense gas elsewhere in SwSt~1.  Therefore, we conclude that the nebula must have an inhomogeneous structure.

We detect nebular absorption at 1040.94 and 1041.69 \AA\ from the two excited fine structure levels of neutral oxygen.  These levels give rise to far-infrared emission lines at 63 and 145 $\micron$ which are often used to infer gas properties, particularly temperature, under the assumption that they are collisionally excited.  We find that the \ion{O}{1} fine structure levels in SwSt~1 have an inverted population ratio.  This requires a non-thermal excitation mechanism, which we identify as fluorescent excitation by the stellar continuum. To the extent that fluorescence affects the level populations, the far-infrared [\ion{O}{1}] line strengths cannot be directly used as diagnostics of density and temperature.

\end{abstract}

\keywords{planetary nebulae: general---planetary nebulae: individual (SwSt~1)---stars: AGB and post-AGB---nucleosynthesis, abundances---ultraviolet: ISM---dust, extinction}

\section{INTRODUCTION}

The planetary nebula (PN) SwSt~1 (Swings \& Struve 1940) is believed to be in the incipience of this stage of evolution, because of its small angular diameter (Kwok et al 1981; Aaquist \& Kwok 1990) and high electron density, found by various studies to have a value in the range $n_{\rm e}$~$\sim$~3$\times$10$^{4}$ to 10$^{5}$~cm$^{-3}$ (Flower et al.\ 1984; de Freitas Pacheco \& Veliz 1987, hereafter dFP87; De Marco et al.\ 2001, hereafter DM01).  The low excitation of its nebular emission line spectrum and cool temperature of the central star, $T_{\rm eff}$~=~40,000~K (DM01) are also suggestive of youth.  The central star of SwSt~1, HD 167362, has been classified as spectral type [WC9] (Crowther et al.\ 1998), a class of H-deficient, C-rich PN central stars that exhibit emission lines indicative of strong winds, although its emission lines are weaker than typical members of this class (Tylenda et al.\ 1993; G\'{o}rny \& Tylenda 2000).  While the evolutionary origin of this class of stars is still unclear, it may be related to late thermal pulses (e.g.\ Herwig et al.\ 1999; Bl\"{o}cker 2001) or binary interactions (De~Marco \& Soker 2002). 

In this paper, we present measurements of nebular absorption lines against the continuum of the central star of SwSt~1.  The \emph{Far Ultraviolet Spectroscopic Explorer} (\emph{FUSE}) spectrum probes material along a narrow ``pencil-beam'' toward the central star, and therefore samples a very small portion of the nebula compared to emission line studies.  The \emph{FUSE} waveband includes the resonance Lyman and Werner bands of molecular hydrogen, several lines of \ion{Fe}{2} and \ion{Fe}{3} (which we use to derive the Fe abundance in SwSt~1), and \ion{Ge}{3}~$\lambda$1088.46, which has been used previously to derive the Ge abundance of five PNe (Sterling et al.\ 2002, hereafter SDB; Sterling \& Dinerstein 2003).  This spectral region also contains several other lines of species in the \ion{H}{2} and neutral atomic regions.

The structure of this paper is as follows:  In \S2, we describe the observations and data reduction.  We discuss the method of analysis in \S3, and present the Voigt component fits used to derive the column densities, central velocities, and Doppler spread parameters of several nebular species. In \S4, we use these results to investigate properties of the ionized gas, such as the electron density and gas phase iron abundance.  We also examine neutral and molecular gas in SwSt~1 (\S5), deriving an upper limit to the column density of nebular molecular hydrogen in the line of sight, and investigating the excitation of the \ion{O}{1} fine structure levels.  The implications of our study are discussed in \S6, and the results are summarized in \S7.

\section{OBSERVATIONS}

The \emph{FUSE} satellite provides access to the wavelength range 905--1187~\AA\ at a nominal resolution $R=15,000$ (FWHM~$=20$~km~s$^{-1}$).  The design (Moos et~al. 2000; Sahnow et~al. 2000) consists of four prime focus telescopes and Rowland-circle spectrographs with microchannel plate detectors.  Two of the channels are coated with Al:LiF, which provides optimal throughput for the spectral range 1000--1187 \AA, while the other two have SiC coatings for optimal reflectivity below 1000 \AA.  Each detector consists of two segments, so that eight channel segments provide coverage of the entire spectral range.  Spectral redundancy ensures that nearly all wavelengths are covered by multiple segments.

SwSt~1 was observed in 2001~August with the large (LWRS, $30\farcs 0 \times 30\farcs 0$) aperture in time-tag mode for a total of 39,587 seconds.  These data are available at the Multi-Mission Archive at the Space Telescope Science Institute (MAST) under the data set identifications B0690101 and B0690102.  We used CALFUSE~(v2.0.5), the standard reduction pipeline available at Johns Hopkins University, to calibrate the raw data.  The exposures were then cross-correlated and co-added in order to increase the signal-to-noise ratio (S/N).  We rebinned the data by three pixels, which improved the average S/N to $\sim$15-25 per rebinned sample, depending on the region of the spectrum.  This increased the effective pixel size from $\sim$~7~m\AA\ to $\sim$~20~m\AA, corresponding to a velocity sampling interval of $\sim$~6.0~km~s$^{-1}$ near 1000~\AA\ and $\sim$~5.2~km~s$^{-1}$ near 1150~\AA.  Figure~1 shows an excerpt from the \emph{FUSE} LiF~2A spectrum of SwSt~1 from 1120--1150~\AA.

We checked for terrestrial airglow contamination of the spectrum (which can be important for lines such as \ion{O}{1}) by comparing the exposures taken during orbital night to the full dataset.  The night spectrum is essentially indistinguishable from that including daytime data, and we therefore conclude that airglow emission has a negligible effect on our results.

We also employ archival \emph{Hubble Space Telescope} (\emph{HST}) ultraviolet observations of SwSt~1, acquired by the Space Telescope Imaging Spectrograph (STIS; Woodgate et~al.\ 1998; Kimble et~al.\ 1998).  This data was obtained in 2000~September, as part of Guest Observer program GO8304 (principal investigator S.\ Kwok).  SwSt~1 was observed for a total of 2300 seconds with the  $0\farcs 2 \times 0\farcs 2$ aperture and E140M echelle grating, which provides a spectral resolution of $R=45,800$ ($FWHM\sim6.5$~km~s$^{-1}$) in the spectral region 1150--1730~\AA.  The data was reduced with the CALSTIS v2.11 processing pipeline (Lindler 1999).  Unfortunately, the E140M spectrum has a low S/N, particularly below 1200~\AA\ (S/N $\leq8$), and therefore we make limited use of this data.

\section{ANALYSIS}

\subsection{Line-Fitting Procedure}

We analyze the \emph{FUSE} data in the segments with the best S/N (i.e.\ largest effective area) for a given wavelength region: LiF~1A for 1000--1083~\AA, and LiF~2A for $\lambda\geq1088$~\AA.  Redundant coverage in the other spectral segments is used to confirm the LiF~1A/2A results.  We normalize each absorption profile by fitting a low-order Legendre polynomial to the continuum and dividing the spectrum by this function.  Atomic wavelengths and data are taken from Morton~(2000; 2003), while H$_2$ wavelengths and oscillator strengths are from the compilation of lines by McCandliss (2003; see also http://www.pha.jhu.edu/\verb+~+stephan/H2ools/h1h2data/, based on Abgrall et~al. 1993a,b).

The line of sight toward the central star of SwSt~1 passes through two interstellar clouds, as well as the absorbing nebular material.  In order to separate the different absorption components of lines, we perform Voigt component fits to each line profile.  These fits yield detailed information about the nebular features, such as central velocity $<v>$, Doppler spread parameter $b$, and column density $N$.  The model profiles are convolved with an instrumental spread function, a Gaussian of FWHM~=~20~km~s$^{-1}$ for \emph{FUSE}, and 6.5~km~s$^{-1}$ for the STIS~E140M data, to account for the systematic broadening effect from the instrumental resolution.

In many cases, more than one line from a given species can be measured.  When possible, we perform simultaneous fits to absorption lines originating in the same level, in order to better constrain the fits and reduce systematic errors.  In a simultaneous fit, the parameters $<v>$, $b$, and $N$ are forced to be the same for all lines from a given level (although the central velocities may differ by a small offset, particularly for lines measured by different instruments).   Simultaneous fitting is especially important for the medium-resolution \emph{FUSE} spectrum, where saturation may not be evident due to instrumental broadening.  We use this technique in several cases to test the importance of saturation effects, and to constrain the parameters of mildly saturated lines which, if modeled individually, would otherwise be highly uncertain.  This procedure is most useful when the fitted lines have significantly different $f$-values, and at least one is not saturated.

We use a different line fitting method in this paper than in previous work for the same dataset (SDB).  While we measure several more nebular lines in SwSt~1 than SDB reported, our two studies have lines of \ion{S}{2}, \ion{S}{3} and \ion{Ge}{3} in common.  By comparing our Table~1 with Table~1 of SDB, one may note that the column densities we measure here are systematically higher than those of SDB, by about 0.3~dex (a factor of two).  This is not due to differences in the atomic data used; indeed, the $f$-values for \ion{S}{3}~$\lambda\lambda1015.78, 1021.32$ are \emph{larger} in Morton (2003) than Morton (1991).  The difference is likely due to the fact that we account for saturation effects in the present study.  We consider the results presented in this paper to be more robust than those of SDB, since we utilize simultaneous fitting, and more recent atomic data (Morton 2003) and solar abundances (Asplund et al.\ 2004).

\subsection{Line Fit Results}

Given the low excitation of the SwSt~1 emission line spectrum (DM01; dFP87), it is not surprising that the only nebular species we detect in the \emph{FUSE} spectrum are neutrals and low ions: \ion{Ar}{1}, \ion{O}{1}, \ion{Fe}{2}, \ion{Fe}{3}, \ion{Ge}{3}, and \ion{S}{3}.  Of these ions, S$^{++}$ requires the highest energy to create (23.3 eV), which is quite modest by PN standards.  There is no evidence of absorption from more highly ionized species.

In Table~1, we show the results of the line fits for all detected nebular species.  Two lines are listed in Column~1 when a simultaneous fit was performed, while one line is recorded otherwise.  The quoted 1-$\sigma$ error estimates include continuum placement and statistical fit uncertainties.  We estimated the statistical errors with Monte Carlo simulations of the absorption line profiles -- 100 artificial spectra were generated for each profile, with the data points modified within the errors of the Gaussian noise.  In most cases, the systematic errors from continuum placement dominate the uncertainties.

For lines arising from different levels of the same ion, we further constrain their fits by forcing the lines to have the same $<v>$ and $b$, although the column densities of the levels are allowed to differ.  This is a useful constraint for ions which exhibit absorption lines from several excited levels (\ion{Fe}{3}, \ion{O}{1}, and \ion{S}{3}), some of which  may be saturated.  Since the excitation energies of the ground term levels for these three ions ($<0.13$~eV) are small compared to the gas temperature (0.9~eV for $T=10,500$~K; DM01), the same value of $b$ should apply for lines from all the observed levels of these ions.

The results of Table~1 are presented graphically in Figures~2a and 2b, where the continuum-normalized profiles are displayed versus heliocentric velocity (all velocities reported in this paper are in the heliocentric frame).  Note that the line of sight toward SwSt~1 intersects two interstellar components, one near $-5$~km~s$^{-1}$, and the other at +20~km~s$^{-1}$ (these are seen most clearly in the profile fits to \ion{Ar}{1} and \ion{Fe}{2}).  These absorbing clouds are well-separated in velocity from nebular absorption.  The profiles of \ion{S}{2}~$\lambda\lambda$1250.58, 1253.81 and \ion{O}{1}~$\lambda\lambda$1304.86, 1306.83 are from STIS data, while all others are from the \emph{FUSE} spectrum.  The total model fit to each line profile (convolved with the instrumental spread function) is shown as a thick, solid line; thin, dashed lines represent individual component fits \textit{before} instrumental broadening.  The depth of the dashed lines therefore indicate the importance of saturation for each component.  In Figures~2a and 2b, the panels are grouped by species -- double panels for simultaneous fits, and single ones for single line fits.  The absorbing ion is indicated above the panel\footnote{Asterisks denote absorption from excited levels; e.g.\ \ion{Fe}{3}$^{**}$ is from the second excited level above the ground state in \ion{Fe}{3}.}, and the wavelength of the line is shown within the panel.  Nearby absorption from other species is indicated in the figures.

The central velocities we measure agree well with the findings of Dinerstein et al.\ (1995b), who used high resolution observations of the \ion{Na}{1}~$\lambda\lambda$5889, 5895 doublet to find a heliocentric velocity of $v_{\rm o}=-47$~km~s$^{-1}$ for the circumstellar gas.  This agreement is somewhat surprising, since ionized species will not necessarily absorb at the same velocity as neutral sodium, which has an ionization potential of only 5.1~eV.  Note that the species with the most aberrant nebular velocities, \ion{Ar}{1} and \ion{O}{1}, are blended with a heavily saturated interstellar component, and $<v>$ is highly uncertain.

\section{RESULTS FOR THE IONIZED REGION}

\subsection{Sulfur Ionic Column Densities}

Sulfur absorption lines are useful for calibrating the abundances of other elements, as well as for determining the average electron density $n_{\rm e}$ along the line of sight.  While the excited levels of the \ion{S}{3} $^3P$ ground term are scantly occupied in the diffuse interstellar medium (ISM), their populations can be significant in the physical conditions of PNe.  Nebular resonance lines are often blended with saturated ISM components, but lines from the excited levels do not suffer from this effect.  The column densities of the excited levels can thus be used to determine the column density of the ground state, under the usual assumption of statistical equilibrium.  Since the nebular components of the resonance transitions of \ion{S}{3} at $\lambda\lambda$1012.50, 1190.20 are lost in the saturated ISM components and blends with other species, we use absorption lines of \ion{S}{3}$^*$ and \ion{S}{3}$^{**}$ to extrapolate the ground state column density, and subsequently derive the total ionic column density.  The populations of these levels are also sensitive to $n_{\rm e}$, and we use the column density ratios to derive the line of sight electron density.

Voigt component fits to \ion{S}{2} and \ion{S}{3} profiles are depicted in Figure~2a, with the fit parameters detailed in Table~1.  We detected absorption from both excited levels of \ion{S}{3}, at $\lambda$1015.78 and $\lambda\lambda$1021.11, 1021.32, in the \emph{FUSE} spectrum.  Since only one \ion{S}{3}$^*$ line provided useful information (the other lines in the \emph{FUSE} and STIS wavebands are too severely blended and/or saturated to significantly constrain the fit), we chose to simultaneously fit \ion{S}{3}$^*$ and \ion{S}{3}$^{**}$, fixing $<v>$ and $b$, but allowing the column densities of the two levels to differ.

The ground term of \ion{S}{2} consists of a single level, so its ionic column density is given by the simultaneous fit of \ion{S}{2}~$\lambda\lambda$1250.58, 1253.81.  \ion{S}{2}~$\lambda$1259.52 is more heavily saturated, and does not significantly constrain the ionic column density when simultaneously fit with the other lines of the multiplet.  The two lines we fit are mildly saturated, and therefore the derived \ion{S}{2} column density is not well determined, but this ion does not contribute a significant amount of the total column of S.

\subsubsection{Electron Density from \ion{S}{3} Level Populations}

In order to extrapolate the ground state population of \ion{S}{3} from those of the excited levels, we used the 5-level task ``nebular.ionic'' by Shaw \& Dufour (1995) in IRAF\footnote{IRAF is distributed by the National Optical Astronomy Observatories, which are operated by the Association of Universities for Research in Astronomy, Inc., under cooperative agreement with the National Science Foundation.}, assuming the electron temperature $T_e=10,500$~K and density $n_{\rm e}=31,600$~cm$^{-3}$ derived from nebular emission lines (DM01).  However, at such a high density, the predicted ratio $R_{12}=$~$N($\ion{S}{3}$^{*})$/$N($\ion{S}{3}$^{**}$)~=~0.96 (hereafter, $R_{\rm ij}$ refers to the population ratio of the $i^{\rm th}$ to the $j^{\rm th}$ levels above the ground state) is nearly a factor of two smaller than the observed value of 1.74$\pm$0.38.  Note that saturation effects in \ion{S}{3}$^*$~$\lambda$1015.78 cannot explain the discrepancy, for if this line were saturated, then $R_{12}$ would be even larger than the observed value.  Furthermore, line-fit uncertainties are unable to bring the observed and predicted line ratios into accord -- forcing the column densities to have a ratio of 0.96 leads to an inferior fit (higher $\chi^2$) for any value of $b$.  We therefore explored the sensitivity of $R_{12}$ to the adopted atomic data and physical parameters.

Using more recent transition probabilities for \ion{S}{3} (Nahar 1993) than the default values for the nebular.ionic task (Mendoza \& Zeippen 1982) has almost no effect on the relative populations of the ground term levels.  However, it should be noted that both sets of values assume LS coupling, which predicts the same relative oscillator strengths of lines within a multiplet.  We also found that varying the electron temperature within a reasonable range (5--15$\times10^3$~K) has little effect on the predicted \ion{S}{3} level populations.  On the other hand, $R_{\rm 12}$ is sensitive to the assumed electron density, and increases as $n_{\rm e}$ decreases.  In Table~2, we show the predicted values of $R_{02}$ and $R_{12}$ for $T_e=10,500$~K and a range of electron densities.  The model agrees best with the observed $R_{12}=1.74\pm0.38$ at $n_{\rm e}=8800$~cm$^{-3}$, with the 1-$\sigma$ errors allowing for the range $n_{\rm e}=6400$--13600~cm$^{-3}$.  It is clear that the observed \ion{S}{3} column density ratio $R_{12}$ is inconsistent with an electron density as high as reported by DM01, $n_{\rm e}=31,600$~cm$^{-3}$.

It is important to note that DM01 derived the electron density from the optical doublet ratio [\ion{S}{2}]~$\lambda$4072/$\lambda$6725.  The $\lambda4072$ lines have high critical densities, and are therefore biased toward dense regions of the nebula.  Other density diagnostics, such as [\ion{S}{2}]~$\lambda6717/\lambda6731$ and [\ion{O}{2}]~$\lambda3726/\lambda3729$, imply $n_{\rm e}\geq10,000$~cm$^{-3}$, but are unreliable since they are near their high density limits.  Therefore, the [\ion{S}{2}]~$\lambda$4072/$\lambda$6725 density indicator is the most reliable one measured by DM01.  Using the same diagnostic ratio, dFP87 also derived a high electron density, $n_{\rm e}=10^5$~cm$^{-3}$.  It is possible that the more highly ionized portion of the nebula has a lower density than that occupied by singly ionized species.  However, this argument is weakened by the fact that Flower et al.\ (1984) derived $n_{\rm e}=10^5$~cm$^{-3}$ from IUE observations of \ion{C}{3}]~$\lambda\lambda1907, 1909$.

Is it possible that the nebular gas along the line of sight to the central star of SwSt~1 has a significantly lower density than other regions of the nebula?  Since the absorption occurs along a ``pencil beam'' sample of the nebular gas in front of the central star, and thus samples a much smaller portion of the nebula than emission line studies, the electron density could be this low if the nebula is inhomogeneous.  Other results (\S\S4.2, 5.1) indicate that this is indeed the case.

We adopt $n_{\rm e}=8800^{+4800}_{-2400}$~cm$^{-3}$ and $T_e=10,500$~K to derive $R_{02}=0.77$ (see Table~2).  This yields $N($\ion{S}{3}~$\lambda$1012.50)~=~0.77$N($\ion{S}{3}$^{**}$)~=~$(6.41\pm2.32)\times10^{14}$~cm$^{-2}$, where the errors include the uncertainties in $n_{\rm e}$.  The adopted column densities for the ground term levels of \ion{S}{3} are summarized in Table~3.

\subsection{Gas Phase Abundance of Iron}

Since the \emph{FUSE} observations do not provide a means of directly determining the overall column density of ionized gas $N$(H$^+$), we derive the Fe abundance in SwSt~1 by using sulfur as a reference element.  S is particularly useful for this purpose, since it is not significantly depleted into dust in the diffuse ISM (e.g.\ Sembach \& Savage 1996; Welty et~al.\ 1999), nor is it affected by nucleosynthesis in the progenitor star.

Because we do not observe all ionization stages of Fe, we use appropriate ionic ratios as surrogates for the elemental abundances.  \ion{Fe}{2} and \ion{Fe}{3} have a similar ionization potential range to \ion{S}{2} and \ion{S}{3} (7.9--30.7 eV for Fe$^+$ and Fe$^{++}$, 10.4--34.8 eV for S$^+$ and S$^{++}$), and therefore we assume that $N($Fe$^+$~+~Fe$^{++}$)/$N($S$^+$~+~S$^{++}$)~$\sim$~(Fe/S).  We use the recommended solar abundances of Asplund et al.\ (2004) to derive the logarithmic abundance ratio relative to solar, [Fe/S]~=~log(Fe/S)$-$log(Fe/S)$_{\odot}$.

As is the case for \ion{S}{3}, the ground terms of \ion{Fe}{2} and \ion{Fe}{3} have multiple fine structure levels that may be significantly populated under nebular conditions.  The column densities of some of these levels could not be measured directly, because absorption lines from the levels were undetected or saturated.  Therefore, we computed the populations for these unobserved levels with models.  We describe the methods used in the following subsections, before presenting the derived Fe abundance.

\subsubsection{\ion{Fe}{2} Level Populations}

Only resonance transitions of \ion{Fe}{2} are detected, despite the fact that this ion has several low-energy levels which may be significantly populated.  The populations of the excited levels must therefore be predicted by models in order to correctly compute the total ionic column density.

Iron is not included in the nebular.ionic task, so we use XSTAR (Kallman \& Bautista 2001) photoionization models to calculate Fe ionic level populations.  XSTAR utilizes atomic data from the Iron Project (Hummer et~al.\ 1993; see Bautista \& Kallman 2001 and references therein), and computes populations of numerous levels for each ion.  We assume a uniform temperature and density nebula, with a blackbody central star of $T_{\rm eff}=40,000$~K, and adopt other model input parameters from Table~12 of DM01.  We tested the sensitivity of the level populations to the adopted parameters by varying elemental abundances, $T_{\rm e}$, radius of the nebula, and central star temperature and luminosity.  The ground term level populations of iron and sulfur ions are negligibly affected by these parameters, as is the derived abundance [Fe/S].

We computed models for several values of $n_{\rm e}$.  From the \ion{S}{3} fine structure levels (\S4.1.1), we adopt $n_{\rm e}=8800$~cm$^{-3}$, with an allowed range $n_{\rm e}=6400$--13600~cm$^{-3}$.  While Fe$^+$ may reside in the photodissociation region as well as the \ion{H}{2} region, we find a relatively small total column density of Fe$^+$ compared to Fe$^{++}$, so that uncertainties in the \ion{Fe}{2} level populations have little effect on the derived Fe abundance.  The resulting Fe level population ratios are listed in Table~3, along with the predicted column densities for unobserved or saturated lines.

We find that a significant fraction of Fe$^+$ resides in excited levels for all of the considered $n_{\rm e}$ (4000--10$^5$~cm$^{-3}$).  The predicted column densities of the excited levels of the \ion{Fe}{2} ground term are consistent with the measured upper limits of lines in the \emph{FUSE} bandpass (Table~1).  To compute the total column density of Fe$^+$, we sum the calculated populations of the thirteen levels of the $^6D$, $^4F$, and $^4D$ terms (higher energy levels contribute less than 1\% to the Fe$^+$ abundance).  The adopted column density for each level is given in Table~3.

\subsubsection{\ion{Fe}{3} Level Populations}

Excited levels in the $^5D$ ground term of \ion{Fe}{3} can give rise to several detectable absorption lines in the \emph{FUSE} bandpass.  \ion{Fe}{3} has no excited terms with energies below 2.4~eV, so it is sufficient to consider only the $^5D$ levels when computing the ionic column density.  We detect lines arising from all ground term levels (Table~1), although the sole line of \ion{Fe}{3}$^{****}$ at $\lambda$1130.40 is quite weak.  The column densities derived from the lines of \ion{Fe}{3} and \ion{Fe}{3}$^*$ are very uncertain due to saturation, particularly that of the resonance line $\lambda$1122.52.  Since a large fraction of \ion{Fe}{3} ions may reside in the two lowest energy levels, we use XSTAR and the absorption lines from the three highest energy $^5D$ levels to estimate their column densities.

The predicted level population ratios are given in Table~3; in general, they correspond poorly to the observed ones.  The discrepancies in $R_{02}$ and $R_{12}$ are due to saturation of the resonance and \ion{Fe}{3}$^*$ lines -- the fitted column densities are lower limits.  Therefore, we adopt the model column densities for \ion{Fe}{3} and \ion{Fe}{3}$^*$.  While $R_{32}$ is in good agreement with the model predictions, $R_{42}$ is too small by a factor of 3-5.  Although \ion{Fe}{3}$^{****}$~$\lambda$1130.40 is only marginally detected, it is very unlikely to be uncertain by such a large amount.  Since $R_{42}$ is not sensitive to the adopted physical parameters of the photoionization models, we examine the atomic data as a source of uncertainty.

It is possible that the oscillator strengths, or $f$-values, of the \ion{Fe}{3} $a^5D-z^5P^0$ multiplet are in error.  While the \ion{Fe}{3} $f$-values of Nahar \& Pradhan (1996, hereafter NP96) are 44\% larger than those recommended by Morton (2003; Ekberg 1993), the \emph{relative} $f$-values within the multiplet are the same in the two studies, since both assume LS coupling and ignore relativistic corrections.  If either of these assumptions does not hold, the relative $f$-values of lines within the multiplet would change.  Alternatively, the adopted transition probabilities (NP96) and collision strengths (Zhang 1996) used by XSTAR to derive the level populations could be uncertain.  However, the ground term level population ratios derived with this data (e.g.\ Keenan et al.\ 2001; Keenan 2004, private communication) agree to within 20\% of those calculated with older atomic data (Keenan et al.\ 1992), showing that our results are not strongly dependent upon the adopted atomic data set.

We are unable to explain the large discrepancy between the observed and predicted $R_{42}$, although \ion{Fe}{3}~$\lambda$1130.40 may have an uncertain $f$-value.  With this caveat in mind, we use the observed $R_{32}$ and $R_{42}$ (which are from unsaturated lines), and the calculated values for $R_{02}$ and $R_{12}$.  The adopted \ion{Fe}{3} column densities are listed in Table~3.

\subsubsection{The Gas Phase Iron Abundance}

We compute the gaseous iron abundance using the sulfur and iron column densities summarized in Table~3, assuming that Fe/S~$\approx~N$(Fe$^+$~+~Fe$^{++}$)/$N$(S$^+$~+~S$^{++}$).  The gaseous Fe abundance is quite high, [Fe/S]~=~$-0.35\pm$0.12, which implies that only about half of the nebular iron is depleted into dust along the line of sight to the central star.  If one uses the \ion{Fe}{3} $f$-values calculated by NP96 rather than those compiled by Morton (2003), [Fe/S] decreases to $-0.50\pm$0.12.

The derived [Fe/S] is shown in Table~4 for the adopted $n_{\rm e}=8800$~cm$^{-3}$ and the 1-$\sigma$ limits, 6400 and 13600~cm$^{-3}$; note that the uncertainties in Table~4 are not the formal errors, since they do not include the uncertainty in $n_{\rm e}$.  For comparison, we also show the Fe abundance derived with the $n_{\rm e}$ value of DM01 ($n_{\rm e}=31,600$~cm$^{-3}$), and of dFP87 and Flower et al.\ (1984; $n_{\rm e}=10^5$~cm$^{-3}$).  The inferred value of [Fe/S] decreases as $n_{\rm e}$ increases, primarily because the \ion{Fe}{3} ground state population relative to excited levels declines as gas density rises.  Furthermore, if we have underestimated the saturation of the \ion{S}{2} lines, the true value of [Fe/S] would be lower (because the actual $N($S$^+$) would be larger than adopted here).  Even given these considerations, the gaseous iron abundance along the line of sight is clearly high compared to the [Fe/H] observed in disk ($-2.2$ to $-1.2$ dex) and even halo ($-0.65$ dex) diffuse interstellar clouds (Sembach \& Savage~1996; Welty et~al.\ 1999).

We stress that the modest level of iron depletion derived from the \emph{FUSE} spectrum is \emph{not} necessarily characteristic of the whole nebula.  In the following subsection, we show that iron is more heavily depleted in other regions of SwSt~1, by deriving the Fe abundance from optical emission lines.

\subsubsection{Comparison to Optical Emission Line Results}

From five optical [\ion{Fe}{3}] lines, dFP87 computed [Fe/H]~=~$-1.56$ and [Fe/S]~=~$-0.88$, using their value of the S abundance.  However, this determination relies on dated atomic data, and therefore may not be reliable.  We re-derive the Fe abundance from optical emission lines, using current atomic data and the optical echelle spectrum of DM01, which was obtained with the Anglo-Australian Telescope (courtesy of O.\ De~Marco; see also Appendix~B of DM01).  We measured line fluxes in the optical spectrum with the ``splot'' routine in IRAF, and de-reddened them using $E(B-V)=0.46$ (DM01).  The resulting line intensities relative to H$\beta$ are reported in Table~5.

We identify several [\ion{Fe}{2}] and [\ion{Fe}{3}] lines, but none from higher ions.  Since most of the optical [\ion{Fe}{2}] emission lines may be severely affected by fluorescence (Rodr\'{i}guez 1999), we derive the Fe$^+$ abundance using only [\ion{Fe}{2}]~$\lambda8617.0$, which is insensitive to fluorescent pumping (Lucy 1995; Baldwin et al.\ 1996).  We identify 12 [\ion{Fe}{3}] emission lines, which are of varying utility (see below) in determining the Fe$^{++}$ abundance.

Several [\ion{Fe}{3}] emission line ratios may be used as nebular diagnostics, as previously discussed by Garstang et al.\ (1978) and Keenan et al.\ (1993).  However, the density diagnostics are often degenerate, allowing for more than one value of $n_{\rm e}$ (Bautista \& Pradhan 1998; Zhang \& Liu 2002).  Using the diagnostic diagrams displayed in Figure~7 of Bautista~\&~Pradhan~(1998), we find that the ratios $I(\lambda4769.4)/I(\lambda4733.9)=0.7\pm0.1$ and $I(\lambda4733.9)/I(\lambda4701.5)=0.5\pm0.1$ indicate either log($n_{\rm e}$/cm$^{-3})=4.6$ or 5.7.  On the other hand, $I(\lambda4754.7)/I(\lambda4733.9)=0.9\pm0.1$ agrees with the other two ratios only for the low log($n_{\rm e}$/cm$^{-3}$) value; the high value significantly exceeds 6.0.  Because of the agreement of all the [\ion{Fe}{3}] density diagnostics at log($n_{\rm e}$/cm$^{-3})=4.6$, and its close concordance to the value derived by DM01, we adopt this value for the electron density.  Unfortunately, none of the temperature-sensitive transitions discussed by Zhang \& Liu~(2002) were detected.  Therefore we adopt $T_{\rm e}=10,500\pm500$~K (DM01) to derive the Fe ionic abundances.

In order to calculate the Fe$^+$ and Fe$^{++}$ abundances (Table~5), we use XSTAR models (see \S4.2.1) to derive the fractional population in the upper level of each transition.  Note that some of the [\ion{Fe}{3}] lines give discrepant ionic abundances compared to other transitions (Table~5).  In particular, [\ion{Fe}{3}]~$\lambda\lambda$4607.0, 4777.7, 4881.0, 4930.5, and 4987.9 indicate significantly different ionic abundances than the other seven lines.  Rodr\'{i}guez (2002) noted that $\lambda\lambda$4607.0 and 4777.7, which yield higher Fe$^{++}$ abundances, may be contaminated by blends with other species; in addition, [\ion{Fe}{3}]~$\lambda$4930.5 may be contaminated by [\ion{O}{3}]~$\lambda$4931.2.  However, blends cannot explain the anomalously low fluxes of $\lambda$4881.0 and $\lambda$4987.9.  Measurement uncertainties are unlikely to be the cause for the inconsistency, particularly in the case of the relatively strong line $\lambda$4881.0.  Interestingly, both of these lines originate in the $a^3H_4$ level.  

In Table~5, we compare the observed line intensities to those predicted by Keenan et al.\ (2001; K01) and Keenan et al.\ (1992; K92, pre-Iron Project) to examine the effects of different sets of atomic data.  While the predicted line intensities are similar in most cases, the K92 values actually reproduce the $a^3H-^5D$ intensities quite well.  By comparison, K01 (who use the same atomic data as we do here) predict $I(\lambda4881.0)$ and $I(\lambda4987.9)$ to be too large by a factor of two or more.  It is possible that the atomic data for the $a^3H-^5D$ multiplet is not reliable, but it should be noted that neither Rodr\'{i}guez (2002) nor Perinotto et al.\ (1999) commented on inconsistencies from these two lines.  In any case, we do not use $\lambda\lambda$4607.0, 4777.7, 4881.0, 4930.5, and 4987.9 in the derivation of the Fe$^{++}$ abundance.

We derive Fe ionic abundances (Table~6) using [\ion{Fe}{2}]~$\lambda$8617.0 and the seven most reliable [\ion{Fe}{3}] lines, assuming log$(n_{\rm e}$/cm$^{-3})=4.6$ and $T_{\rm e}=10,500$~K.  We vary log$(n_{\rm e}$/cm$^{-3})$ from 4.5 to 5.0, in accord with previously derived electron densities (DM01; dFP87; Flower et al.\ 1984), and assume an uncertainty of $\pm500$~K for $T_{\rm e}$, as given by DM01, to produce the error bars on the ionic abundances listed in Table~6.  As discussed above, Fe$^+$ is a trace ion in the nebula, so the derived Fe abundance is not strongly affected even if much of the Fe$^+$ resides in the photodissociation region, under significantly different physical conditions than we assume.

In order to correct for unseen stages of ionization (primarily Fe$^{+3}$), we adopt the ionization correction factor (ICF) given by Rodr\'{i}guez~(2002):
\[ \rm Fe/H = 1.1\times(\rm O/O^+)\times(\rm Fe^{++}/\rm H^++\rm Fe^+/\rm H^+). \]
Using data from Tables~10 and 11 of DM01 (we adopt the ``empirical abundance'' for O in their Table~11), we find ICF(Fe)~$=1.1\times(\rm O/\rm O^+)=1.16\pm0.56$.  The relatively large error is due to the uncertainty in the total O abundance.  This results in [Fe/H]$=-1.64\pm0.24$, or [Fe/S]$=-1.15\pm0.33$, using the S abundance from DM01.  In other words, the Fe abundance we derive from the optical data of DM01 agrees well with that of dFP87.  Iron is depleted by a factor of 10 or more according to the optical data, which starkly contrasts with the very modest depletion we find from the \emph{FUSE} data.

Thus, it appears that the high gaseous Fe abundance that characterizes the column toward the central star is not representative of the emitting ionized gas.  This result is strengthened by the fact that the [Fe/S] derived from the \emph{FUSE} spectrum is not strongly dependent upon the assumed $n_{\rm e}$ (Table~4).

\subsection{Abundance of the \emph{s}-process Element Ge}

The slight nebular Fe depletion along the line of sight implies that other refractory elements will also be relatively undepleted.  SDB derived the Ge abundance in four PNe (including SwSt~1) from the resonance line \ion{Ge}{3}~$\lambda$1088.46 in the \emph{FUSE} band, and reported a range of possible Ge abundances, depending on the unknown degree of depletion.  The range found for SwSt~1 spanned [Ge/S]~=~(0.86--1.48)$\pm0.15$ (note the typographical error in Table~2 of SDB), depending on whether Ge is wholly gaseous, or is depleted to the same extent as in cold interstellar clouds ($-0.62$~dex; Savage \& Sembach 1996).  The modest Fe depletion indicates that [Ge/S] is at the low end of the range found by SDB.

We have re-evaluated the Ge abundance with our new method of analysis (\S3.1) and updated atomic data from Morton (2003).  Using the ionization correction factor Ge/S~= $N$(Ge$^{++}$)/$N$(S$^+$~+~S$^{++}$), based on similarities of the  ionization potential ranges, we find [Ge/S]~=~$0.72\pm$0.06.  This agrees within the errors with the value [Ge/S]~=~$0.86\pm$0.15 calculated by SDB.

Many theoretical and observational studies of the \emph{s}-process discuss \emph{n}-capture element yields relative to iron.  Assuming that Ge/Fe~=~$N$(Ge$^{++}$)/$N$(Fe$^{++}$), we derive [Ge/Fe]~=~$1.08\pm0.13$ using \ion{Fe}{3} $f$-values from Morton (2003), or $1.24\pm0.13$ using values from NP96.  (Note that [Ge/Fe] is not equivalent to [Ge/S]~--~[Fe/S]; the ionization potential range of Fe$^+$ does not coincide with that of Ge$^{++}$, and is therefore not used to derive [Ge/Fe]).  In Table~4, we display the Ge abundance relative to S and Fe for the range of densities implied by UV \ion{S}{3} absorption lines, as well as the $n_{\rm e}$ found from optical emission lines.  If iron is depleted, then the derived value of [Ge/Fe] is an upper limit.  However, [Ge/S] is enhanced by at least a factor of 5 for the entire range of electron densities we consider.

\section{RESULTS FOR THE NEUTRAL/MOLECULAR REGION}

\subsection{Non-Detection of Nebular H$_2$}

The original motivation for {\it FUSE} GO program B069, for which this observation of SwSt~1 was taken, was to study the molecular components of PNe through examination of the H$_2$ Lyman and Werner resonance lines seen in absorption against the continua of their central stars. Accordingly, targets were selected to have velocities well-separated from low-velocity foreground interstellar material, and to display near-infrared quadrupole emission lines from vibrationally-excited H$_2$ or some other indicator of neutral or molecular material. SwSt~1 was selected as a target for the H$_2$ search because it meets both of these criteria. 

\subsubsection{Previous Evidence for Neutral/Molecular Material in SwSt~1}

SwSt~1 was found by Dinerstein et al.\ (1995b; hereafter DSU) to possess nebular components in the \ion{Na}{1} ``D'' lines at 5890.0, 5895.9 \AA.  Since neutral Na has a very low ionization potential (5.1 eV) and appears to correlate with other indicators of atomic and molecular material (see Table 5 of DSU), \ion{Na}{1} is presumed to be a tracer of neutral material in nebulae. 

In the case of pure absorption by foreground material, the \ion{Na}{1} lines can in principle be used to infer a column density of circumstellar material from the classical ``doublet ratio'' method  (e.g. Dinerstein \& Sneden 1988), which has been widely applied to diffuse interstellar clouds.  However, this simple approach is probably invalid for SwSt~1, where the modest equivalent widths of the \ion{Na}{1} lines (55$\pm$15~m\AA) appear inconsistent with their ratio, which is close to the fully saturated value of unity.  The reason for this inconsistency is evident from Figure~5 of DSU, where it can be seen that the nebular absorption in SwSt~1 is partially filled in by scattered \ion{Na}{1} photons; the true absorption profiles are actually deeper than they appear.    The observed absorption minima fall at a heliocentric radial velocity of --47~km~s$^{-1}$, blueshifted from the systemic emission line value of --20~km~s$^{-1}$; the blending of the emission and absorption components is not likely to affect the apparent velocity by more than a few km~s$^{-1}$.

Circumnebular \ion{H}{1} 21~cm absorption was subsequently reported by Gussie \& Taylor (1995), at a velocity of $\sim$--38~km~s$^{-1}$.  They did not, however, detect CO $J$ = 2--1 millimeter emission, although all of the other four PNe they detected in \ion{H}{1} were also seen in CO. Huggins et al.\ (1996) also searched for CO $J$ = 2--1 in SwSt~1 and did not detect it.  This suggests that SwSt~1 possesses only a modest amount of neutral/molecular material.

SwSt~1 is a source of near-infrared H$_2$ emission, although its spectrum has not been studied as intensively as some other PNe. In their Appendix B, DM01 reported a detection of the 2.122~$\micron$ $v$ = 1--0 S(1) line at a strength about 5\%\ that of \ion{H}{1} Br$\gamma$.  Additional H$_2$ lines, in particular $v$ = 1--0 S(0)~2.223~$\micron$ and several $Q$-branch lines, are visible in the spectrum of DM01, although the line fluxes are not cited in their Table B2.  Likkel et al.\ (2004) also detect 2.122 and 2.223 $\micron$, but no other H$_2$ lines; their data do not extend longward of 2.26 $\micron$ and therefore do not include the $Q$-branch.  They measure a 2.223/2.122 $\micron$ flux ratio consistent with radiative excitation of moderately dense gas in a strong radiation field. If radiative excitation is occurring, the Lyman and Werner lines in the \emph{FUSE} band, which initiate the fluorescent cascades, should be detectable as long as there is H$_2$ along the absorption column.
 
\subsubsection{Upper Limits on $N(H_2)$ for Specific $J$ Levels}

We do not detect nebular H$_2$ absorption in the \emph{FUSE} spectrum of SwSt~1.  In order to quantify this statement, we measure limits on the equivalent widths of nebular components of specific H$_2$ transitions on the blue wings of the corresponding interstellar features.  Because the velocity shift is modest (only 2--3 spectral resolution elements), this is not feasible for lines that are very saturated in the ISM, namely those arising from the $J$~=~0 and 1 levels. Instead, we examine lines from $J$ $\ge$ 2, which have less saturated and hence narrower interstellar components.  We select several lines, arising from each of the $J$~=~2, 3, 4, and 5 levels, that are free from interference by other species and fall in regions with good S/N ratio in the continuum.

The selected transitions are listed in Table 7, and we display a representative sample of these lines in Figure~3.  For comparison, we show $J=0$ and 1 lines in the upper panels, from which it can be seen that blending with the interstellar component (thick solid line) prevents meaningful measurements of limits on the nebular column density.  We use two methods to set upper limits on the equivalent widths of the nebular components.  In some cases, we were able to ``fit'' a nebular feature by interpolating a conservatively high continuum near the nebular velocity.  For an alternate estimate, we evaluated the RMS noise level in the adjacent continuum, assumed a line width equal to the spectral resolution element, and took the corresponding 3-$\sigma$ value as our upper limit.  The $W_{\lambda}$ values from both methods, typically $\sim$~5--15 m\AA, were converted to column densities $N_{\rm J}$ by assuming that the lines are on the linear portion of the curve of growth.  The $N_{\rm J}$ limits are given in Columns 4 and 5 of Table 7, and depicted in Figure~3 by dotted lines showing Voigt profiles with the measured $N_{\rm J}$ limit at the nebular velocity, assuming $b=11.4$~km~s$^{-1}$ (the value measured for nebular \ion{O}{1} lines).  Overall, our upper limits are approximately $N_{\rm J}\sim5\times10^{13}$~cm$^{-2}$ for $J$~=~2 and 3, and $N_{\rm J}\sim4\times10^{13}$~cm$^{-2}$ for $J$~=~4 and 5. 

\subsubsection{Upper Limits on the Total H$_2$ Column Density}

In order to translate these level-specific column densities into a total H$_2$ column, it is necessary to make assumptions about the excitation equilibrium.  Given $f_{\rm J}$, the fraction of H$_2$ molecules in level $J$, then it is possible to compute $N_{\rm tot}$ = $N_{\rm J}$/$f_{\rm J}$ for each $J$. For purposes of a rough estimate, we assume that the lowest few $J$ levels are dominated by collisional processes, so that their populations are predicted by Boltzmann statistics. To normalize the level population $N_{\rm J}$ = $g_{\rm J}~e^{-(\chi_J/kT)}$, we need the appropriate partition function $Z(T)$ = $\Sigma$~($N_{\rm J}$) over all $J$ levels that contain significant populations.

We compute partition functions for H$_2$ for several representative values of the rotation temperature $T_{\rm rot}$: 200, 500, 800, and 1600~K. We include the 20 energy levels with the largest populations, which consist of the $J$ = 0 -- 15 levels of the $v$ = 0 vibrational state, and $J$ = 0 -- 5 for $v$ = 1, which begins to interleave the energy level ladder of $v$ = 0 between $J$~=~8 and 9.  For radiative excitation, a single value of $T_{\rm rot}$ does not strictly apply; however, the effective values of $T_{\rm rot}$ for $J$~$\le$~5 for moderately high radiation field and gas densities lie in the range $\sim$~200 -- 500~K (Draine 2004, private communication).  Furthermore, in BD+30$^{\circ}$3639, which has highly excited circumstellar H$_2$, $T_{\rm rot}$~$\sim$~800~K for $v$~=~0, while the $v~\ge$~2 levels have substantially higher values of $T_{\rm rot}$ (Dinerstein \& Bowers 2004; Dinerstein et al.\ 2005, in preparation). 

In Table 8 we convert the upper limits on the specific $J$ levels into the corresponding total column densities $N_{\rm tot}$(H$_2$).  In general, the strongest constraints on the total amount of H$_2$ come from the $J$ levels with the largest fractional populations, but the $J$ value with the largest fraction varies with temperature.  For $T_{\rm rot}$~$\ge$~500~$K$ the tightest constraint comes from the $J$~=~3 lines: $N_{\rm tot}$(H$_2$)~$\sim$~2~$\times$~10$^{14}$~cm$^{-2}$.  For the lowest temperature we calculate here (more appropriate for lower-density gas and a weak radiation field), the best limit, $N_{\rm tot}(H_2)~\le$~7~$\times$~10$^{14}$~cm$^{-2}$, is from $J$~=~2.

It should be noted that the derived upper limit to $N_{\rm tot}$(H$_2)$ may be in error if the H$_2$ is sufficiently cold that only the $J$~=~0 and 1 states are significantly populated, since we were not able to derive meaningful upper limits to the column densities of these levels.  For example, if $T_{\rm rot}=100$~K, then 1.2\% of the H$_2$ would be in the $J=2$ state, so that our upper limit of $N_{\rm tot}$(H$_2)=N_2/f_2~\le$~7~$\times$~10$^{14}$~cm$^{-2}$ (derived for $T_{\rm rot}=200$~K)  would be too low by a factor of about 6.  Temperatures lower than 60~K would be necessary for our estimate to be too low by two orders of magnitude.  However, given that H$_2$ in the diffuse ISM, far from strong UV radiation fields, exhibits rotational temperatures of $\geq$80--90~K (Shull et al.\ 2004; Savage et al.\ 1977), it is very unlikely that circumstellar material close to the luminous, hot central star of a PN would be this cold.

\subsubsection{Comparison to H$_2$ Seen in Emission}

The flux of H$_2$~1-0~S(1)~2.122~$\micron$ was reported as $(1.5\pm0.2)\times$~10$^{-13}$~erg~cm$^{-2}$~s$^{-1}$ by DM01, and $(1.1\pm0.3)\times$~10$^{-13}$~erg~cm$^{-2}$~s$^{-1}$ by Likkel et al.\ (2004).  While there are no direct determinations of the size of the H$_2$-emitting region, the similarity of the fluxes measured in different aperture sizes implies that it must be relatively compact.  \emph{HST} images of the ionized gas suggest a knotty ring of approximate dimensions 1\farcs3~$\times$~0\farcs9, or 4.3~$\times$~10$^{-11}$ ster (DM01). Since in most PNe the H$_2$-emitting regions are slightly more extended than the ionized gas (e.g. Latter et al.\ 1995), we assume that the H$_2$ fills solid angle $\sim$~5~$\times$~10$^{-11}$ ster. This corresponds to a surface brightness for the 2.122 $\micron$ line of (2.6~$\pm$~0.4)~$\times$~10$^{-3}$~erg~cm$^{-2}$~s$^{-1}$~ster$^{-1}$.  Even this large value, 20 times the surface brightness of the ``canonical'' fluorescent H$_2$ PN Hubble 12 (Dinerstein et al.\ 1988; Ramsay et al.\ 1993), is only a lower limit, since it is beam-averaged and the emission may not be uniform.

For optically thin emission, this surface brightness implies a beam-averaged column density in the originating level of the transition ($v$~=~1,~$J$~=~3) of 1.0~$\times$~10$^{17}$~cm$^{-2}$.  Even in a highly excited cloud, this level contains only a small fraction of the total H$_2$: for thermal excitation, $f_{1,3}$(800$K$)~$\sim$~2$\times$10$^{-4}$ and $f_{1,3}$(1600$K$)~$\sim$~7$\times$10$^{-3}$.  Therefore the total column density of H$_2$ would have to be at least $N_{\rm tot}$(H$_2$)~$\gtrsim$~10$^{19}$~cm$^{-2}$.  Comparing this estimate to our limits on the line of sight $N_{\rm tot}$(H$_2$) from the UV data, we see that they differ by \emph{at least} four orders of magnitude.  If the absorption line of sight contained the column of H$_2$ indicated by the infrared emission lines, the nebular components of the UV lines would be deeply saturated, even if the H$_2$ is very hot ($T_{\rm rot}=1600$~K; dashed lines in Figure~3).  This disparity indicates an extreme asymmetry or clumpiness in the distribution of molecular material in SwSt~1.

\subsection{Excitation of \ion{O}{1} in SwSt~1}

\subsubsection{UV Absorption Lines from Excited \ion{O}{1}} 

Despite the absence of nebular H$_2$ absorption in SwSt~1, some neutral species are clearly present in the \emph{FUSE} spectrum.  Some of the low-ionization species we detect, such as \ion{Ar}{1}, \ion{Fe}{2}, and \ion{S}{2}, may exist in both the ionized nebula and in the transition region between ionized and neutral material (the photodissociation region, or PDR).  However, we also detect absorption from \ion{O}{1}, which has an ionization potential nearly identical to that of \ion{H}{1}, and therefore is expected to reside primarily in the neutral zone. 

Under the conditions of a PDR, the excited fine structure levels of the \ion{O}{1} $^{3}P$ ground term may be significantly populated, and therefore can be detected in absorption along with resonance transitions.  If the excitation of these levels is exclusively due to collisions, one can use their column density ratios as a measure of the kinetic gas temperature.  While the lines of the well-known resonance triplet at 1302-1306~\AA\ in the STIS waveband suffer from saturation and blending, the \emph{FUSE} band offers a variety of alternate \ion{O}{1} multiplets, with a range of different $f$-values and hence sensitivities to saturation effects.  Many of these lie below 1000~\AA, where the spectrum is crowded and has low S/N, so we focus on the multiplet near 1040~\AA.  Although the ground state line ($\lambda$1039.23) cannot be used due to saturation and blending with the ISM component, simultaneous fits show that the absorption lines from the excited levels in this multiplet are free from such effects (see Figure~2a).

Taking the ratio of the column densities derived from these fits (Table 1), we find $N$(\ion{O}{1}$^{**}~\lambda1041.69$)/$N$(\ion{O}{1}$^{*}~\lambda1040.94)=0.71\pm$0.16.  This value exceeds by about a factor of two the ratio of statistical weights (1/3), which is the limiting value for LTE in the infinite temperature case, $T$~$>>$~$\Delta\chi/k$ (where $\Delta\chi$ is the energy difference between the two levels).  This population inversion demonstrates incontrovertibly that collisions are not the primary excitation mechanism for the fine structure levels of \ion{O}{1} in SwSt~1.

\subsubsection{Alternate Excitation Mechanisms} 

Several processes other than collisional excitation can populate the excited levels of \ion{O}{1}. These processes were discussed and evaluated in detail by Grandi (1975a, b), in order to understand the strength of \ion{O}{1} $\lambda$8446.5 and other permitted lines of \ion{O}{1} in the Orion nebula.  These excitation mechanisms include recombination into excited levels; fluorescent excitation by the stellar continuum; or fluorescent excitation by \ion{H}{1} Lyman lines, primarily Ly~$\beta$ but also to some extent Ly~$\epsilon$ and Ly~$\zeta$.  In order to distinguish among these possibilities, we examine the intensities of various optical and near-infrared \ion{O}{1} lines in SwSt~1 as reported in the literature.
 
We use the data of DM01 (see \S4.2.4 for more details) to measure optical and near-infrared \ion{O}{1} lines.  The resulting ratios of various \ion{O}{1} line fluxes to $\lambda$8446.5 are summarized in Table 9. We also list the corresponding ratios in Orion (Esteban et al.\ 2004), for which Grandi (1975a, b) concluded that fluorescent excitation by the stellar continuum was the only mechanism that could explain the observed pattern of lines.

Recombination can be ruled out as the dominant \ion{O}{1} excitation mechanism in SwSt~1 by virtue of the strength of \ion{O}{1}~$\lambda8446.5$ relative to H$\alpha$: the line flux ratio is $\sim$0.003, over a factor of 30 larger than predicted for recombination (Rudy et~al.\ 1989). Additional support comes from the weakness of \ion{O}{1}~$\lambda$7774.2 relative to $\lambda$8446.5; if recombination populated these levels, these two lines should have comparable strengths.  Starlight excitation is usually cited as the mechanism which produces \ion{O}{1}~$\lambda$8446.5 in Galactic nebulae (Grandi 1976), but Rudy et al.\ (1989) have argued, on the basis of measurements of near-infrared \ion{O}{1} lines, that Ly~$\beta$ fluorescence is responsible for exciting \ion{O}{1} in the compact PN IC~4997. However, Lyman line fluorescence cannot explain the strengths of \ion{O}{1} lines from states other than $^{3}D$, such as $\lambda\lambda$5554.9, 6046.4, and 13164 (Grandi 1975a; Rudy et al.\ 1989), which cannot be populated via line excitation. 

The general consistency between the \ion{O}{1} line intensities in SwSt~1 and Orion seen in Table 9 strongly supports the conclusion that the excitation of the higher levels of \ion{O}{1} is dominated by stellar continuum fluorescence. Since the atoms will cascade to lower levels, eventually reaching the fine structure levels of the ground $^{3}P$ term, we suggest that the population inversion we find in the \emph{FUSE} spectrum is a result of this fluorescent cascade initiated by stellar continuum photons.

\subsubsection{Implications for the Interpretation of the Far-Infrared [\ion{O}{1}] Emission Lines in PNe}

[\ion{O}{1}]~$^{3}P_{1}$~--~$^{3}P_{2}$~63~$\micron$, the transition from the first excited fine structure level to the ground state, is one of the strongest cooling lines in PDRs.  Although [\ion{C}{2}] 158 $\micron$ is the strongest line at low densities, the 63~$\micron$ line dominates the cooling of PDRs with intense incident UV fields and high densities, conditions that apply to the circumstellar PDRs of PNe (Tielens \& Hollenbach 1985; Hollenbach \& Tielens 1997).  The companion [\ion{O}{1}] $^{3}P_{0}$~--~$^{3}P_{1}$ line at 145~$\micron$ is weaker than 63~$\micron$ due to the higher energy and lower statistical weight of its originating level. Under the assumption that collisional excitation dominates the level populations, ratios among the three lines can be used as diagnostics of the local thermal parameters $n$ and $T$ (e.g. Watson 1984). 

However, the populations in the \ion{O}{1} ground term can be redistributed by fluorescent excitation in a strong stellar radiation field.  As seen above, stellar continuum fluorescence can overpopulate the upper level of the [\ion{O}{1}]~145~$\micron$ line relative to its value from purely collisional excitation.  Therefore, when starlight excitation affects the ground term levels of \ion{O}{1}, the true PDR temperature will be lower than indicated by the line flux ratios of [\ion{O}{1}]~63, 145, and [\ion{C}{2}]~158~$\micron$.  In the gas column of SwSt~1 probed by \emph{FUSE}, starlight fluorescent excitation is so dominant that there is no ambiguity about its importance. However, in other PDRs, more modest contributions might be present but not easily recognized.  

Observations with NASA's \emph{Kuiper Airborne Observatory} (\emph{KAO}) and the \emph{Infrared Space Observatory (ISO)} have been used to determine the physical parameters of PDRs for several PNe, although SwSt~1 itself has not been observed.  In order to interpret the line ratios, a correction must be made for the elevated C/O abundances of these objects.  Applying such corrections to measurements from the \emph{KAO}, Dinerstein et al. (1995a, 1997) found log($n$/cm$^{-3})\sim4.3$, $T\ge1000$~K for the PDRs associated with BD+30$^{\circ}$3639, IC~418, NGC~6572, and NGC~7027.  (We cite only a lower limit for $T$, since the ratios become insensitive to the precise value for $T>1000$~K.)  It is interesting to note that \ion{O}{1} optical and near-infrared line ratios indicate that stellar continuum fluorescence is important in \textit{all four} of these PNe (BD+30$^{\circ}$3639 -- Aller \& Hyung 1995, Rudy et al.\ 1991b; IC~418 -- Sharpee et al.\ 2003; NGC~6572 -- Hyung et al.\ 1994, Rudy et al.\ 1991a; NGC~7027 -- Baluteau et al.\ 1995, Rudy et al.\ 1992).  Liu et al. (2001) observed the same four objects as well as 18 additional PNe, with the Long Wavelength Spectrometer on \emph{ISO}.  With the exception of NGC~7027, the measured flux ratios of Liu~et~al.\ differ by up to a factor of two from those found by Dinerstein et al., yielding lower temperatures ($T\sim200$--400~K) and much higher densities (log($n$/cm$^{-3}\sim5.3-5.5$) for these PNe.  The reason for this difference is not clear.  One possible source of the discrepancy is the correction for C/O, which is uncertain by as much as a factor of 2--3 in PNe.  Another factor is that the \emph{ISO} observations were taken with a larger entrance aperture than the \emph{KAO} data, and may include a greater contribution from more extended, cooler gas, where the stellar radiation field is weaker and collisional excitation dominates.  Thus, the contribution of the fluorescently excited component may be diluted compared to that seen by \emph{KAO}.

This effect has probably been seen, albeit not recognized, in other sources such as the $\rho$~Oph cloud (Liseau et al. 1999a) and Carina nebula (Mizutani et al.\ 2004), where excess emission is seen in the 145~$\micron$ line compared to the predictions of PDR models and other diagnostics of gas temperature. Indeed, Liseau et al. (1999a) specifically suggest that an unidentified nonthermal pumping mechanism may be responsible (also see Liseau et al.\ 1999b). Our \emph{FUSE} observations of SwSt~1 provide an explanation for this widespread puzzle.

\section{DISCUSSION}

\subsection{Nature of the Inhomogeneities in SwSt~1}

Many of the physical properties that we derive from the \emph{FUSE} spectrum of SwSt~1 -- the electron density, the high gaseous iron abundance, and the absence of nebular H$_2$ absorption -- differ dramatically from values derived previously from emission lines.  However, \emph{FUSE} probes only a ``pencil-beam'' of gas in the line of sight to the central star, a small portion of the nebula compared to emission line studies; the differences therefore suggest that the absorption line of sight is not representative of conditions elsewhere in the nebula.  These results indicate a non-uniform distribution of ionized and molecular material in SwSt~1, either due to a highly asymmetrical global structure, or extreme clumping.

It is possible that global structure can explain the observations.  The images of the optical emission in DM01 suggest a toroidal morphology, tipped with respect to our line of sight. (Assuming that the optical nebula is really a circular ring, the observed aspect ratio implies an inclination of $\sim$60\degr\ from pole-on.)  In this case, the \emph{FUSE} spectrum looks through a polar cone in which the gas has mostly been evacuated. The line of sight to the central star, which traverses this cone, would then be a region deficient in both molecules and dust, where the gas density is low compared to the toroidal ring.

Alternatively, small-scale clumping may be responsible for the nebular inhomogeneities.  In fact, such clumpiness may be quite common in PNe.  For example, O'Dell et al.\ (2002) find numerous knots in nearby PNe, and Gon\c{c}alves et al.\ (2001) discuss 50 PNe with detected low ionization structures.  Small-scale structure may form quite early in the evolution of a PN, from instabilities created during the interaction of the fast and slow winds (Dwarkadas \& Balick 1998) or with the passage of the ionization front (e.g.\ Garc\'{i}a-Segura \& Franco 1996).  Condensations in the progenitor's envelope during the AGB phase could also contribute to knots in the ejecta (Dyson et~al.\ 1989), although Huggins \& Mauron (2002) find a lack of evidence that this occurs.

Recent studies have shown that PNe with H-deficient central stars (such as SwSt~1) may be particularly likely to form clumpy nebulae.  Acker et al.\ (2002) and Pe\~na et al.\ (2003) find that PNe with [WC] central stars tend to exhibit a higher degree of turbulence (which implies the presence of small-scale instabilities) than those with H-rich stars.  Indeed, the high metal abundance and mass-loss rates of [WC] central stars likely prolong the momentum-conserving phase of the nebula (Acker et~al.\ 2002), which is prone to nonlinear thin shell instabilities (Dwarkadas \& Balick 1998).  Furthermore, clumpy winds from [WC] central stars, as detected by Grosdidier et al.\ (2000),  can create or enhance inhomogeneities in the nebula.  While Grosdidier et~al.\ did not detect inhomogeneous outflows from the central star of SwSt~1, they acknowledge that their observations were insufficient to detect wind variability on timescales longer than about an hour.  However, they detected clumpy winds from the central stars of NGC~40 and BD+30$^{\circ}$3639, which are similar to that of SwSt~1.

Since the angular scale of SwSt~1 is so small that even \emph{HST} images cannot fully resolve its intrinsic structure, it is not possible to assess whether the non-uniform distribution of nebular material is due to global asymmetries or clumping.  However, it is clear that the nebula is inhomogeneous, and the sight line to the central star does not intersect material with the properties exhibited by emission lines.

\subsection{Inhomogeneities of the Dust and Molecular Gas in SwSt~1}

The spatial segregation of H$_2$ into dense regions, implied by the absence of UV H$_2$ absorption in an object with near-IR H$_2$ emission, is not unexpected in view of the fragility of H$_2$ molecules to photodissociation and the intensity of the stellar UV radiation field.  Tielens (1993) estimated that the lifetime of H$_2$ molecules outside of dense knots -- where self shielding and dust extinction protect them from photodissociation -- is on the order of a few years, and therefore predicted that H$_2$ in PNe must be confined to dense clumps.  Recent high spatial resolution observations of H$_2$ and CO emission in the Helix and Ring Nebulae, where small linear dimensions can be resolved, have confirmed this expectation (Huggins et al.\ 2002; Speck et al.\ 2003).  Even given the relative youth of SwSt~1 (estimated as $\sim$~300 yr by DM01), the timescales for the propagation of photodissociation fronts are sufficiently short (e.g.\ Hollenbach \& Natta 1995) that the H$_2$ will be largely dissociated along paths which do not contain condensations with $N>10^{19-20}$~cm$^{-2}$.

The origin of the inhomogeneities in the gas phase Fe abundance and the corresponding dust-to-gas ratio in SwSt~1 is less clear.  Our results show a difference of $\sim1$ dex between the gas phase Fe abundance in the UV absorption column and the emitting gas, where it is depleted by about a factor of $\sim40$, similar to values found in the warm ionized ISM and \ion{H}{2} regions (Sembach \& Savage 1996; Welty et al.\ 1999; Rodr\'{i}guez 2002).  The low optical Fe$^{++}$ abundance is corroborated by Likkel et al.\ (2004), who find  similarly low Fe$^{++}$/H$^+$ from near-infrared [\ion{Fe}{3}] lines in SwSt~1.  Further evidence for the presence of dust in SwSt~1 comes from the detection of thermal dust emission in the infrared.  SwSt~1 exhibits the dual dust chemistry that is typical of PNe with [WC] central stars (Cohen et~al.\ 2002), as indicated by the presence of features from both O-rich (Aitken et~al.\ 1979; Zhang \& Kwok 1990) and C-rich dust (Roche et~al.\ 1996; Szczerba et al.\ 2001).  The O-rich grains dominate by mass: silicates account for about 95\% of the dust, with the other 5\% giving rise to the aromatic emission bands usually attributed to PAHs (Casassus et al.\ 2001).  On general considerations, Fe should be strongly depleted in regions with abundant silicate grains (Draine 2003), and has also been found to be underabundant in nebulae with C-rich dust (e.g. Shields 1978;\ Perinotto et al.\ 1999).

There are two possible scenarios that could produce the observed difference in Fe depletion between the absorption column and the emitting regions. If the ejected material initially had a uniform dust-to-gas ratio, then the dust must have been destroyed more efficiently in lower density regions of the nebula.  Alternatively, variations in dust formation or growth efficiency in a clumpy outflow could have caused a non-uniform dust-to-gas ratio, and the initial inhomogeneity remains imprinted on the evolving nebula.

The main dust destruction agents in nebulae are shocks and energetic stellar photons.  Although the central star of SwSt~1 drives a fast wind (900~km~s$^{-1}$; DM01), the wind-shocked gas should lie in the central cavity, and be highly ionized and heated to $T\sim10^{5-6}$~K (e.g.\ Chu et al.\ 2004).  Shocks moving into the dense optically emitting material would be much slower (Tielens 1993), and therefore less able to destroy the dust.  The fact that we measure radial velocities for the UV absorption lines that are consistent with the overall expansion velocity of the optical emitting gas (DM01) argues that the absorbing gas was not accelerated by a shock.  In addition, the ionization balance seen from the UV absorption lines is consistent with that indicated by the emission lines.  Therefore, based on kinematic and excitation arguments, shock processing of the gas and dust in SwSt~1 seems improbable.

Unlike H$_2$ molecules, dust grains are relatively robust with respect to evaporation by stellar radiation, with the exception of small grains (containing fewer than 20--30 atoms) that have small heat capacities relative to the energies of typical stellar UV photons.  These small grains cannot re-radiate the absorbed photon energy via an equilibrium process, and can undergo large temperature fluctuations.  Such grains, whether carbon or silicate-based, can be heated to temperatures exceeding the sublimation temperature of refractory elements after absorbing just one or two high-energy photons (e.g.\ Guhathakurta \& Draine 1989).

Observational studies have drawn mixed conclusions about radiative destruction of dust in PNe.  Observed gas phase depletions of elements such as Fe and Mg do not show any clear trends with nebular age (Middlemass 1988; Perinotto et al.\ 1999).  Similarly, it does not appear that dust-to-gas ratios derived from infrared emission are lower in older, more evolved PNe (Hoare et al.\ 1992).  Such arguments depend on comparing different objects, which may not be appropriate if the initial dust formation efficiency differs among their progenitor stars.  Another approach is to search for evidence of dust destruction within specific nebulae. In this regard, the fact that the dust-to-gas ratios derived for the ionized and neutral (PDR) regions in NGC~7027 agree to within a factor of two is taken as evidence against dust destruction in the ionized gas by Hoare et al.\ (1992). On the other hand, there is evidence for a ``depletion gradient'' in some PNe in the sense that some refractory elements (e.g.\ Al, Ca, and Mg) are less depleted in regions of higher than those of lower ionization, suggesting that atoms have been released from grains into the gas phase by the stellar radiation field (Kingdon et al.\ 1995; Kingdon \& Ferland 1997).

In SwSt~1, we find very disparate gas phase Fe depletions in different regions of the nebula \emph{with the same ionization states}, which makes it seem unlikely that there has been a substantial difference in the amount of radiative dust destruction between these two regions.  This indicates that the disparate Fe depletions are most likely caused by initial inhomogeneities in the dust formation or growth efficiency in the AGB outflow.  Such inhomogeneities could be caused by low temperature structures (e.g.\ starspots) in the atmosphere of the progenitor AGB star, as suggested by Frank (1995) and Soker \& Clayton (1999).  In addition, the grain growth efficiency is higher in denser regions of a clumpy outflow (e.g.\ Dorfi \& H\"{o}fner 1996), so that the grain size distribution would be skewed toward larger radii in dense gas.  If the density inhomogeneities are preserved into the PN phase, dust in the more rarefied gas is more susceptible to photodestruction because of the smaller grain size distribution.  This implies that radiative dust destruction could have been more efficient in lower density regions of SwSt~1, even though the incident radiation field is similar for the low and high density gas (as indicated by the comparable ionization balance of the absorbing and emitting gas).

The dust deficit in the gas probed by the UV absorption lines may have a direct influence on the amount of H$_2$ present, since the primary channel for H$_2$ formation is through grain surface reactions.  The removal of this mechanism for reforming H$_2$ is likely to lower the equilibrium H$_2$ abundance that can be maintained in the presence of a dissociating and ionizing radiation field (Aleman \& Gruenwald 2004 and references therein).  Therefore, although we attribute the extreme contrast in the H$_2$ abundance primarily to photodestruction, it is possible that the dust inhomogeneities amplify and maintain the spatial variations in molecular material.  

\subsection{Nucleosynthesis and Dredge-Up in the Progenitor Star}

Germanium ($Z=32$) isotopes can be produced by slow neutron(\emph{n})-capture nucleosynthesis (the \emph{s}-process) in PN progenitor stars during the AGB phase of evolution.  Nuclei created by the \emph{s}-process may be convectively transported to the stellar envelope along with carbon during third dredge-up (TDU; e.g.\ Busso et~al.\ 1999).  The fact that we find an enhanced Ge abundance relative to solar in SwSt~1 implies that the \emph{s}-process and TDU were efficient in the progenitor star.

We measure a Ge enrichment [Ge/S]~=~$0.72\pm$0.06, or [Ge/Fe]~=~$1.08\pm0.13$, which can be compared with theoretical models of stellar \emph{s}-process enrichments, given the metallicity.  The S abundance derived by DM01 and dFP87 implies that the progenitor star of SwSt~1 may have had a metallicity as low as $\sim0.3Z_{\odot}$.  However, the O and Ne abundances indicate that the metallicity is closer to solar.  The derived Ge overabundance is slightly larger than predicted by recent theoretical models of the \emph{s}-process in this metallicity range (Busso et~al. 2001; Busso 2003, private communication; Goriely \& Mowlavi 2000).  However, many uncertainties exist in the strength of TDU and in the effect of convective overshoot and rotation on the \emph{s}-process (e.g.\ Herwig et~al.\ 2003; Siess et~al.\ 2004).  Therefore, we consider the derived Ge enrichment to be consistent with evolutionary models, within uncertainties in the metallicity and yields.

\section{CONCLUSIONS}

We now summarize the principal results of our study:

1.  We have analyzed the \emph{FUSE} spectrum of nebular gas in front of the central star of the compact PN SwSt~1.  The absorbing nebular gas exhibits a low degree of ionization, consistent with previous studies of this object.  We find that the absorbing species in SwSt~1 lie at a heliocentric velocity of $-40$ to $-50$~km~s$^{-1}$, in good agreement with Na~D measurements.

2.  From the column density ratio of \ion{S}{3} fine structure levels, we derive an electron density $n_{\rm e}=8800^{+4800}_{-2400}$~cm$^{-3}$, which is at least a factor of three lower than found in previous optical and UV emission line studies.

3.  We derive the abundance of iron relative to sulfur, accounting for populations of all of the low-energy levels of the relevant ions.  We find [Fe/S]~=~$-0.35\pm$0.12 or $-0.50\pm0.12$, depending on the adopted $f$-values.  The gaseous iron abundance is insensitive to uncertainties in $n_{\rm e}$ and $T_{\rm e}$, and is high in comparison to interstellar medium values, and to the Fe abundance derived from optical [\ion{Fe}{3}] emission lines in SwSt~1 ([Fe/H]$=-1.64\pm0.24$ and [Fe/S]$=-1.15\pm0.33$).  This indicates the presence of marked inhomogeneities in the dust-to-gas ratio in different regions of the nebula.

4.  The high gaseous Fe abundance indicates that other refractory elements should not be strongly depleted in the sight line.  Thus, the elemental Ge abundance lies at the low end of the range suggested by SDB.  We re-derive [Ge/S] and find a value of 0.72$\pm$0.06.  The large Ge overabundance implies that the progenitor of SwSt~1 experienced \emph{s}-process nucleosynthesis and efficient third dredge-up while on the AGB.

5.  We do not detect nebular H$_2$ absorption in the \emph{FUSE} spectrum.  We estimate an upper limit to the column density of $N($H$_2)<7\times10^{14}$~cm$^{-2}$.  This limit is more than four orders of magnitude smaller than the beam-averaged column density derived from near-infrared vibrationally excited H$_2$ lines.  Apparently, little of this molecular material lies along the line of sight to the central star.

6.  The low electron density, non-detection of nebular H$_2$, and high gaseous Fe abundance derived from the \emph{FUSE} spectrum of SwSt~1 implies that the gas lying in the direction of the central star has markedly different properties from the regions of the nebula sampled by emission lines.  Since the \emph{FUSE} spectrum samples only the gas along the line of sight toward the central star, this indicates that both the ionized and molecular gas in SwSt~1 are non-uniformly distributed.  However, it is not possible to determine whether the nebular inhomogeneities arise from global asymmetries or small-scale clumping of nebular material, since SwSt~1's angular size is small enough that its internal structure is not resolved in \emph{HST} images.

7.  SwSt~1 exhibits absorption features from excited fine structure levels of neutral oxygen at 1040.94 and 1041.69 \AA.  These levels must have been populated by a non-collisional process, as indicated by the inverted population structure.  Using optical \ion{O}{1} line ratios, we find that fluorescent pumping by the stellar continuum is the most likely excitation mechanism.  The fine structure levels of the \ion{O}{1} ground term give rise to lines at 63 and 145~$\mu$m, whose diagnostic value is compromised if the level populations are affected by fluorescent processes, rather than being determined purely by collisional excitation.

The \emph{HST} data presented in this paper were obtained from the Multimission Archive at the Space Telescope Science Institute (MAST). STScI is operated by the Association of Universities for Research in Astronomy, Inc., under NASA contract NAS 5-26555. Support for MAST for non-HST data is provided by the NASA Office of Space Science via grant NAG 5-7584 and by other grants and contracts.  We are indebted to O.\ De~Marco for making the optical spectrum of SwSt~1 available to us, and S.\ McCandliss for providing H$_2$ line data.  We also thank F.\ Keenan, who supplied us with \ion{Fe}{3} level population data, and M.\ Barlow, B.\ Draine, and S.\ Federman for helpful suggestions and discussion.  T.\ Kallman provided invaluable assistance with the operation of the XSTAR code.  This work was supported by NASA contracts NAG~5-9239, NAG~5-11597, and NSF grant AST~97-31156.

\begin{deluxetable}{llccc}
\tablecolumns{5}
\tablewidth{0pc} 
\tablecaption{SwSt 1 Nebular Component Fits}
\tablehead{
\colhead{} & \colhead{$\lambda$} & \colhead{$<v>$} & \colhead{$b $} & \colhead{$N$} \\
\colhead{Ion} & \colhead{(\AA)} & \colhead{(km s$^{-1}$)} & \colhead{(km s$^{-1}$)} & \colhead{(cm$^{-2}$)}}
\startdata
\ion{Ar}{1} & 1048.22, 1066.66 & $-57.6\pm$1.6\tablenotemark{c} & 2.4$\pm$0.5 & 9.12$\pm$7.89 (14)\tablenotemark{d}\\
%\ion{Fe}{2} & 1125.45 & $-54.5\pm$1.2 & 8.5$^{+1.1}_{-6.3}$ & 1.86$\pm$0.57 (14)\\
\ion{Fe}{2} & 1143.23, 1144.94 & $-47.0\pm$1.3 & 7.9$\pm$2.5 & 5.37$\pm$0.74 (13)\\
\ion{Fe}{2}$^{*}$ & 1148.28 & \nodata & \nodata & $<$ 1.51 (13)\\
\ion{Fe}{2}$^{**}$ & 1150.69 & \nodata & \nodata & $<$ 1.55 (13)\\
\ion{Fe}{2}$^{***}$ & 1148.96 & \nodata & \nodata & $<$ 1.17 (13)\\
\ion{Fe}{2}$^{****}$ & 1154.40 & \nodata & \nodata & $<$ 1.00 (13)\\
\ion{Fe}{3} & 1122.52\tablenotemark{b} & $-40.5\pm$1.7 & 7.0$^{+0.4}_{-1.8}$ & 6.76$^{+43.36}_{-1.75}$ (14)\tablenotemark{d}\\
\ion{Fe}{3}$^{*}$ & 1124.87, 1128.04\tablenotemark{b} & $-40.5\pm$1.7 & 7.0$^{+0.4}_{-1.8}$ & 2.09$^{+1.07}_{-0.18}$ (14)\tablenotemark{d}\\
\ion{Fe}{3}$^{**}$ & 1128.72, 1131.91\tablenotemark{b} & $-40.5\pm$1.7 & 7.0$^{+0.4}_{-1.8}$ & 2.00$\pm$0.32 (14) \\
\ion{Fe}{3}$^{***}$ & 1129.19, 1131.19\tablenotemark{b} & $-40.5\pm$1.7 & 7.0$^{+0.4}_{-1.8}$ & 1.38$\pm$0.19 (14)\\
\ion{Fe}{3}$^{****}$ & 1130.40\tablenotemark{b} & $-40.5\pm$1.7 & 7.0$^{+0.4}_{-1.8}$ & 1.55$\pm$0.25 (13)\\
\ion{Ge}{3} & 1088.46 & $-48.0\pm$0.5 & 7.2$\pm$1.8 & 5.75$\pm$0.66 (12)\\
\ion{O}{1} & 1039.23 & $-58.2\pm$1.5\tablenotemark{c} & 7.6$\pm$2.8 & 1.62$\pm$0.57 (15)\tablenotemark{d}\\
\ion{O}{1}$^{*}$ & 1040.94, 1304.86\tablenotemark{a,b} & $-43.2\pm$3.3 & 11.4$^{+2.7}_{-0.5}$ & 1.12$\pm$0.18 (15)\\
\ion{O}{1}$^{**}$ & 1041.69, 1306.03\tablenotemark{a,b} & $-43.2\pm$3.3 & 11.4$^{+2.7}_{-0.5}$ & 7.94$\pm$1.29 (14)\\
\ion{S}{2} & 1250.58\tablenotemark{a}~, 1253.81\tablenotemark{a} & $-43.3\pm$2.8 & 6.7$\pm$1.2 & 1.07$\pm$0.12 (14)\\
\ion{S}{3}$^{*}$ & 1015.78\tablenotemark{b} & $-46.1\pm$0.7 & 6.1$\pm$0.5 & 1.45$\pm$0.24 (15)\\
\ion{S}{3}$^{**}$ & 1021.11, 1021.32\tablenotemark{b} & $-46.1\pm$0.7 & 6.1$\pm$0.5 & 8.32$\pm$1.16 (14)\\
\enddata
\tablecomments{The resulting central heliocentric velocity, Doppler spread parameter, and column density are given for the nebular component of the indicated species.  The fits are displayed as solid lines in Figures~2a and 2b.  The uncertainties listed are 1-$\sigma$ estimates, and upper limits are 3-$\sigma$.  Note that the notation 1.51~(13)~=~$1.51\times10^{13}$.}
\tablenotetext{a}{The parameters of these lines are measured from the STIS spectrum.}
\tablenotetext{b}{Parameters obtained by simultaneously fitting lines from different levels of the ion, fixing $<v>$ and $b$ but allowing the column densities of the levels to differ.}
\tablenotetext{c}{Highly uncertain $<v>$ due to heavily saturated interstellar components.}
\tablenotetext{d}{Uncertain column density due to saturation and/or blending.  The measured column density is not used in abundance measurements.}
\end{deluxetable}

\clearpage

\begin{deluxetable}{lcc}
\tablecolumns{3}
\tablewidth{300.0pt} 
\tablecaption{\ion{S}{3} Column Density Ratios}
\tablehead{
\colhead{$n_{\rm e}$} & \colhead{} & \colhead{}\\
\colhead{(cm~$^{-3}$)} & \colhead{$R_{02}$\tablenotemark{a}} & \colhead{$R_{12}$\tablenotemark{a}}}
\startdata
4000 & 1.83 & 2.87\\
5000 & 1.42 & 2.49\\
6000 & 1.14 & 2.21\\
7000 & 0.96 & 2.01\\
8000 & 0.85 & 1.85\\
8800\tablenotemark{b} & 0.77 & 1.74\\
9000 & 0.77 & 1.71\\
10000 & 0.70 & 1.63\\
15000 & 0.51 & 1.31\\
20000 & 0.41 & 1.14\\
25000 & 0.36 & 1.04\\
31600\tablenotemark{c} & 0.32 & 0.96\\[0.15in]
Observed & \nodata & 1.74$\pm$0.38\\
\enddata
\tablecomments{All ratios are computed for an electron temperature $T_e=10,500$~K and the indicated electron density, using Shaw \& Dufour's (1995) nebular.ionic task in IRAF.  The adopted transition probabilities are from Nahar (1993), while the collision strengths are from Galavis et al.\ (1995).}
\tablenotetext{a}{$R_{02}=N($\ion{S}{3})/$N($\ion{S}{3}$^{**}$) and $R_{12}=$~$N($\ion{S}{3}$^{*})$/$N($\ion{S}{3}$^{**}$).}
\tablenotetext{b}{Adopted $n_{\rm e}$.  Observational (1-$\sigma$) errors allow for a range of $n_{\rm e}$ from 6400--13600~cm$^{-3}$.}
\tablenotetext{c}{Electron density derived by DM01.  Note the poor agreement with the observed $R_{12}$.}
\end{deluxetable}

\clearpage

\begin{deluxetable}{lcccccc}
\tablecolumns{7}
\tablewidth{0pt}
\tablecaption{Adopted Sulfur and Iron Level Column Densities}
\tablehead{
\colhead{} & \colhead{Index} & \colhead{} & \colhead{} & \colhead{$R_{\rm ij}$} & \colhead{Observed} & \colhead{Adopted}\\
\colhead{Ion} & \colhead{$i$\tablenotemark{a}} & \colhead{Level} & \colhead{$R_{\rm ij}$\tablenotemark{b}} & \colhead{Range\tablenotemark{c}} & \colhead{$R_{\rm ij}$} & \colhead{$N$ (cm$^{-2}$)\tablenotemark{d}}}
\startdata
\ion{S}{3} & 0 & $^3P_0$ & 0.77 & 0.53--1.05 & \nodata & 6.41$\pm$2.32 (14)\\
 & 1 & $^3P_1$ & 1.74 & 1.36--2.12 & 1.74$\pm$0.38 & 1.45$\pm$0.24 (15)\tablenotemark{e}\\
 & 2 & $^3P_2$ & 1.00 & 1.00--1.00 & 1.00 & 8.32$\pm$1.16 (14)\tablenotemark{e}\\
\ion{Fe}{2} & 0 & $a~^6D_{9/2}$ & 1.00 & 1.00--1.00 & 1.00 & 5.37$\pm$0.74 (13)\tablenotemark{e}\\
 & 1 & $a~^6D_{7/2}$ & 0.24 & 0.16--0.32 & $<$~0.28 & 1.31$\pm$0.36 (13)\\
 & 2 & $a~^6D_{5/2}$ & 0.14 & 0.10--0.19 & $<$~0.29 & 7.35$\pm$2.44 (12)\\
 & 3 & $a~^6D_{3/2}$ & 8.8 (-2) & 6.4--12.3 (-2) & $<$~0.22 & 4.73$\pm$1.58 (12)\\
 & 4 & $a~^6D_{1/2}$ & 4.7 (-2) & 3.5--6.4 (-2) & $<$~0.19 & 2.54$\pm$0.77 (12)\\
 & 5 & $a~^4F_{9/2}$ & 1.02 & 0.86--1.08 & \nodata & 5.50$\pm$0.97 (13)\\
 & 6 & $a~^4F_{7/2}$ & 0.15 & 0.10--0.22 & \nodata & 7.84$\pm$3.24 (12)\\
 & 7 & $a~^4F_{5/2}$ & 7.4 (-2) & 4.7--11.7 (-2) & \nodata & 3.98$\pm$1.86 (12)\\
 & 8 & $a~^4F_{3/2}$ & 4.9 (-2) & 3.2--7.7 (-2) & \nodata & 2.65$\pm$1.22 (12)\\
 & 9 & $a~^4D_{7/2}$ & 2.7 (-2) & 1.9--4.2 (-2) & \nodata & 1.46$\pm$0.60 (12)\\
 & 10 & $a~^4D_{5/2}$ & 9.2 (-3) & 5.6--16.7 (-3) & \nodata & 4.93$\pm$2.99 (11)\\
 & 11 & $a~^4D_{3/2}$ & 4.8 (-3) & 2.8--9.3 (-3) & \nodata & 2.58$\pm$1.74 (11)\\
 & 12 & $a~^4D_{1/2}$ & 2.4 (-3) & 1.4--4.6 (-3) & \nodata & 1.26$\pm$0.86 (11)\\
\ion{Fe}{3} & 0 & $^5D_4$ & 14.1 & 9.9--18.7 & $>$~3.4\tablenotemark{f} & 2.82$\pm$0.99 (15)\\
 & 1 & $^5D_3$ & 1.64 & 1.61--1.64 & 1.05$^{+0.56}_{-0.19}$\tablenotemark{f} & 3.28$\pm$0.52 (14)\\
 & 2 & $^5D_2$ & 1.00 & 1.00--1.00 & 1.00 & 2.00$\pm$0.32 (14)\tablenotemark{e}\\
 & 3 & $^5D_1$ & 0.73 & 0.70--0.76 & 0.69$\pm$0.15 & 1.38$\pm$0.19 (14)\tablenotemark{e}\\
 & 4 & $^5D_0$ & 0.38 & 0.32--0.45 & 7.8$\pm$1.8 (-2) & 1.55$\pm$0.25 (13)\tablenotemark{e}\\
\enddata
\tablenotetext{a}{The index $i$ denotes the excitation of the absorbing species, e.g.\ $i=2$ for \ion{Fe}{3}$^{**}$. }
\tablenotetext{b}{The level population ratio of level~$i$ relative to \ion{S}{3}$^{**}$, \ion{Fe}{2}, or \ion{Fe}{3}$^{**}$, for an electron density of $n_{\rm e}=8800$~cm$^{-3}$.}
\tablenotetext{c}{The range in $R_{\rm ij}$ when $n_{\rm e}$ is varied from 6400--13600~cm$^{-3}$.}
\tablenotetext{d}{Errors for predicted column densities include the uncertainty in $n_{\rm e}$, as well as those from the measurement of the reference line.}
\tablenotetext{e}{The observed column density is adopted for the abundance calculation.}
\tablenotetext{f}{Saturated line.}
\end{deluxetable}

\clearpage

\begin{deluxetable}{lccc}
\tablecolumns{4}
\tablewidth{0pt} 
\tablecaption{Dependence of Fe and Ge Abundances on Electron Density}
\tablehead{
\colhead{$n_{\rm e}$} & \colhead{} & \colhead{} & \colhead{}\\
\colhead{(cm$^{-3}$)} & \colhead{[Fe/S]\tablenotemark{a}} & \colhead{[Ge/S]} & \colhead{[Ge/Fe]\tablenotemark{a}}}
\startdata
6400 & $-0.28\pm$0.02 & 0.69$\pm$0.05 & 0.98$\pm$0.05\\
8800\tablenotemark{b} & $-0.35\pm$0.02 & 0.72$\pm$0.05 & 1.08$\pm$0.05\\
13600 & $-0.44\pm$0.02 & 0.74$\pm$0.05 & 1.21$\pm$0.05\\
31600 & $-0.58\pm$0.02 & 0.76$\pm$0.05 & 1.39$\pm$0.05\\
10$^5$ & $-0.68\pm$0.02 & 0.77$\pm$0.05 & 1.53$\pm$0.05\\
\enddata
\tablecomments{Logarithmic abundances relative to solar (Asplund et al.\ 2004) are given, for a range of electron densities $n_{\rm e}$.  The listed 1-$\sigma$ uncertainties are not the formal errors, since they do not account for uncertainties in $n_{\rm e}$.}
\tablenotetext{a}{Assumes $f$-values from Morton (2003).  If NP96 $f$-values are used for \ion{Fe}{3}, [Fe/S] decreases by 0.15 dex, and [Ge/Fe] increases by 0.16 dex.}
\tablenotetext{b}{The adopted electron density and abundances.}
\end{deluxetable}

\clearpage

\begin{deluxetable}{lcccccc}
\tablecolumns{7}
\tablewidth{0pt} 
\tablecaption{Optical Iron Emission Line Intensities}
\tablehead{
\colhead{} & \colhead{$\lambda_{\rm lab}$} & \colhead{$\lambda_{\rm obs}$} & \colhead{$10^3I_{\lambda}$/} & \colhead{Predicted $10^3I_{\lambda}/I_{\rm H\beta}$} & \colhead{Predicted $10^3I_{\lambda}/I_{\rm H\beta}$} & \colhead{}\\
\colhead{Ion} & \colhead{(\AA)} & \colhead{(\AA)} & \colhead{$I_{\rm H\beta}$} & \colhead{K01\tablenotemark{a} ; log $n_{\rm e}=4.5$} & \colhead{K92\tablenotemark{a} ; log $n_{\rm e}=5.0$} & \colhead{(Fe$^{+i}$/H$^+$)\tablenotemark{b}}}
\startdata
$[$\ion{Fe}{2}$]$ & 8617.0 & 8615.9 & 0.5$\pm$0.1 & \nodata & \nodata & 2.0 (-8)\\
$[$\ion{Fe}{3}$]$ & 4607.0\tablenotemark{c} & 4606.6 & 1.5$\pm$0.1 & 1.0 & 0.8 & 8.0 (-7)\\
 & 4658.1 & 4657.6 & 12.3$\pm$0.7 & 12.3 & 12.3 & 5.5 (-7)\\
 & 4701.5 & 4701.0 & 5.1$\pm$0.2 & 5.3 & 5.3 & 5.2 (-7)\\
 & 4733.9 & 4733.3 & 2.6$\pm$0.2 & 2.5 & 2.7 & 5.6 (-7)\\
 & 4754.7 & 4754.2 & 2.2$\pm$0.2 & 2.3 & 2.2 & 5.3 (-7)\\
 & 4769.4 & 4768.9 & 1.8$\pm$0.2 & 1.8 & 1.7 & 5.3 (-7)\\
 & 4777.7\tablenotemark{c} & 4777.1 & 1.7$\pm$0.1 & 1.2 & 1.3 & 7.5 (-7)\\
 & 4881.0 & 4880.4 & 4.2$\pm$0.2 & 10.5 & 4.4 & 2.1 (-7)\\
 & 4930.5\tablenotemark{c} & 4930.1 & 1.1$\pm$0.2 & 0.7 & \nodata & 8.5 (-7)\\
 & 4987.9 & 4986.6 & 1.0$\pm$0.1 & 2.0 & 0.8 & 2.5 (-7)\\
 & 5084.8 & 5084.2 & 0.4$\pm$0.1 & 0.4 & 0.3 & 5.1 (-7)\\
 & 5270.4 & 5269.9 & 6.3$\pm$0.3 & 6.5 & 4.0 & 5.5 (-7)\\
\enddata
\tablecomments{Optical Fe line intensities, corrected for reddening using $E(B-V)=0.46$ (DM01), are given relative to H$\beta$.  The optical data were obtained by DM01 from the Anglo-Australian Telescope in 1993 May, and were subsequently reduced by those authors.  The listed 1-$\sigma$ errors account only for statistical uncertainties in the line fits.}
\tablenotetext{a}{Predicted [\ion{Fe}{3}] emission line intensities (scaled to $\lambda$4658.1) from K01~=~Keenan et al.\ (2001) and K92~=~Keenan et al.\ (1992), for $T=10^4$~K and the $n_{\rm e}$ that best agrees with the observed intensities.  See text for further discussion.}
\tablenotetext{b}{The ionic abundance derived from the given line with $T_{\rm e}=10,500$~K and log($n_{\rm e}/$cm$^{-3}$)=4.6.  Transition probabilities are from NP96, and collision strengths from Zhang~(1996).  The adopted ionic abundances are given in Table~6.}
\tablenotetext{c}{Blended line.}
\end{deluxetable}

\clearpage

\begin{deluxetable}{lr}
\tablecolumns{7}
\tablewidth{300.0pt} 
\tablecaption{Iron Abundance From Optical Emission Lines}
\tablehead{
\colhead{Abundance} & \colhead{}\\
\colhead{or ICF} & \colhead{Result}}
\startdata
Fe$^+$/H$^+$ & 2.0$\pm$0.3 (-8) \\
Fe$^{++}$/H$^+$ & 5.3$\pm$0.4 (-7) \\
ICF\tablenotemark{a} & 1.16$\pm$0.56 \\
Fe/H & 6.39$\pm$3.13 (-7) \\
$[$Fe/H$]$ & $-1.64\pm$0.24 \\
$[$Fe/S$]$\tablenotemark{b} & $-1.15\pm$0.33 \\
\enddata
\tablecomments{The adopted ionic and total iron abundances are displayed.  The errors incorporate uncertainties in $n_{\rm e}$, $T_{\rm e}$, and the line fits.  The ionization correction factor (ICF), and hence total Fe abundance, allow for the large uncertainties in O/O$^+$ quoted in DM01.}
\tablenotetext{a}{The ICF is taken from Rodr\'{i}guez (2002): Fe/H$=1.1\times($O/O$^+)\times($Fe$^{++}$/H$^++$Fe$^+$/H$^+$).}
\tablenotetext{b}{[Fe/S] is derived using the S abundance from DM01.}
\end{deluxetable}

\clearpage

\begin{deluxetable}{lcccc}
\tablecolumns{5}
\tablewidth{0pc} 
\tablecaption{Column Density Upper Limits for H$_2$ Lyman Lines}
\tablehead{
\colhead{Line} & \colhead{$\lambda$ (\AA)} & \colhead{$f$} & \colhead{$N$(Fit)\tablenotemark{b}} & \colhead{$N$(RMS)\tablenotemark{c}}}
\startdata
8-0 R(2) & 1003.98 & 1.662 (-2) & \nodata & $<$ 7.89 (13)\\
8-0 P(2) & 1005.39 & 9.911 (-3) & $<$ 7.59 (13) & $<$ 1.40 (14)\\
0-0 Q(2)\tablenotemark{a} & 1010.94 & 2.381 (-2) & \nodata & $<$ 4.99 (13)\\
7-0 P(2) & 1016.46 & 1.023 (-2) & \nodata & $<$ 9.44 (13)\\
4-0 P(2) & 1053.28 & 9.021 (-3) & \nodata & $<$ 1.05 (14)\\
2-0 P(2) & 1081.27 & 4.699 (-3) & \nodata & $<$ 1.95 (14)\\
1-0 P(2) & 1096.44 & 2.366 (-3) & $<$ 4.57 (14) & $<$ 3.55 (14)\\
\tableline
7-0 R(3) & 1017.42 & 1.838 (-2) & $<$ 5.37 (13) & $<$ 6.07 (13)\\
4-0 R(3) & 1053.98 & 1.336 (-2) & $<$ 8.32 (13) & $<$ 5.61 (13)\\
4-0 P(3) & 1056.47 & 9.556 (-3) & $<$ 6.46 (13) & $<$ 8.37 (13)\\
3-0 P(3) & 1070.14 & 7.537 (-3) & \nodata & $<$ 1.24 (14)\\
2-0 R(3) & 1081.71 & 6.363 (-3) & $<$ 2.82 (13) & $<$ 1.43 (14)\\
1-0 P(3) & 1099.79 & 2.526 (-3) & \nodata & $<$ 2.90 (14)\\
\tableline
5-0 R(4) & 1044.54 & 1.548 (-2) & $<$ 5.50 (13) & $<$ 4.77 (13)\\
4-0 R(4) & 1057.38 & 1.294 (-2) & $<$ 3.55 (13) & $<$ 7.49 (13)\\
3-0 P(4) & 1074.31 & 7.744 (-3) & \nodata & $<$ 9.73 (13)\\
2-0 P(4) & 1088.80 & 5.151 (-3) & $<$ 1.23 (14) & $<$ 1.98 (14)\\
\tableline
0-0 Q(5)\tablenotemark{a} & 1017.83 & 2.389 (-2) & \nodata & $<$ 4.39 (13)\\
3-0 R(5) & 1075.24 & 9.279 (-3) & \nodata & $<$ 9.39 (13)\\
\enddata
\tablenotetext{a}{Line from the Werner band.}
\tablenotetext{b}{Upper limit derived from fit within the noise at the nebular velocity.}
\tablenotetext{c}{Upper limit derived from RMS noise in the adjacent continuum.}
\end{deluxetable}

\clearpage

\begin{deluxetable}{cccccccccc}
\tablecolumns{10}
\tablewidth{0pt} 
\tabletypesize{\scriptsize}
\tablecaption{Estimated $N_{\rm tot}$(H$_2$) Upper Limits for Various Values of $T_{\rm rot}$}
\tablehead{
\colhead{$J$} & \colhead{$N_{\rm J}$\tablenotemark{a}} & \colhead{$f_{\rm J}(200$ K)\tablenotemark{b}} & \colhead{$f_{\rm J}(500$ K)\tablenotemark{b}} & \colhead{$f_{\rm J}(800$ K)\tablenotemark{b}} &  \colhead{$f_{\rm J}(1600$ K)\tablenotemark{b}} & \colhead{$N_{\rm tot}(200$ K)\tablenotemark{c}} & \colhead{$N_{\rm tot}(500$ K)\tablenotemark{c}} & \colhead{$N_{\rm tot}(800$ K)\tablenotemark{c}} & \colhead{$N_{\rm tot}(1600$ K)\tablenotemark{c}}  \\
\colhead{} & \colhead{(cm$^{-2}$)} & \colhead{} & \colhead{} & \colhead{} & \colhead{} & \colhead{(cm$^{-2}$)} & \colhead{(cm$^{-2}$)} & \colhead{(cm$^{-2}$)} & \colhead{(cm$^{-2}$)}}
\startdata
2 & $<5.0$ (13) & 7.3 (-2) & 1.4 (-1) & 1.3 (-1) & 9.1 (-2) & 6.8 (14) & 3.5 (14) & 3.7 (14) & 5.5 (14)\\
3 & $<5.0$ (13) & 2.4 (-2) & 2.2 (-1) & 3.0 (-1) & 2.8 (-1) & 2.1 (15) & 2.3 (14) & 1.6 (14) & 1.8 (14)\\
4 & $<4.0$ (13) & 4.0 (-4) & 2.5 (-2) & 5.6 (-2) & 7.8 (-2) & 1.0 (17) & 1.6 (15) & 7.1 (14) & 5.1 (14)\\
5 & $<4.0$ (13) & 2.0 (-5) & 1.8 (-2) & 7.3 (-2) & 1.7 (-1) & 2.0 (18) & 2.2 (15) & 5.5 (14) & 2.3 (14)\\
\enddata
\normalsize
\tablenotetext{a}{The H$_2$ column density in rotational level $J$, from the upper limits in Table~7.}
\tablenotetext{b}{The fraction $f_{\rm J}$ of total H$_2$ molecules in level $J$, for a given temperature.}
\tablenotetext{c}{Total H$_2$ column density for a given rotational temperature.}
\end{deluxetable}

\clearpage

\begin{deluxetable}{lcc}
\tablecolumns{3}
\tablewidth{0pt} 
\tablecaption{Optical \ion{O}{1} Emission Line Ratios}
\tablehead{
\colhead{} & \colhead{Observed} & \colhead{Orion}\\
\colhead{Line} & \colhead{$I(\lambda)/I(\lambda8446.5)$} & \colhead{$I(\lambda)/I(\lambda8446.5)$\tablenotemark{a}}}
\startdata
\ion{O}{1}]~$\lambda4368.2$ & 1.0$\pm$0.2 (-1) & 8.3 (-2)\\
\ion{O}{1}]~$\lambda5512.8$ & 4.9$\pm$0.5 (-2) & 2.7 (-2)\\
\ion{O}{1}]~$\lambda5554.9$ & 4.2$\pm$0.8 (-2) & 2.8 (-2)\\
\ion{O}{1}]~$\lambda5958.5$ & 8.1$\pm$0.8 (-2) & 4.3 (-2)\\
\ion{O}{1}~$\lambda6046.4$ & 1.5$\pm$0.1 (-1) & 1.0 (-1)\\
\ion{O}{1}~$\lambda7002.1$ & 5.6$\pm$0.4 (-2) & 9.8 (-2)\\
\ion{O}{1}~$\lambda7254.4$ & 4.0$\pm$0.5 (-2)\tablenotemark{b} & 1.2 (-1)\\
\ion{O}{1}~$\lambda7774.2$ & 3.3$\pm$0.5 (-2) & \nodata\\
\ion{O}{1}~$\lambda13164$ & 0.3$\pm$0.1 & \nodata \\
\ion{H}{1}~$\lambda4861.3$ & 129 & 113\\
\ion{H}{1}~$\lambda6562.8$ & 102 & 326\\
\enddata
\tablecomments{Optical line ratios of \ion{O}{1}, H$\alpha$, and H$\beta$ are given relative to \ion{O}{1}~$\lambda$8446.5, where $I(\lambda8446.5)=(2.5\pm0.1)\times10^{-13}$ erg~cm$^{-2}$~s$^{-1}$~\AA$^{-1}$.  The optical data were obtained by DM01 from the Anglo-Australian Telescope in 1993 May, and were subsequently reduced by those authors.  The 1-$\sigma$ errors account for statistical uncertainties only.}
\tablenotetext{a}{Orion emission line ratios from Esteban et al.\ (2004).  These ratios are representative of stellar continuum-excited fluorescence.}
\tablenotetext{b}{Uncertain -- contaminated by telluric feature.}
\end{deluxetable}

\clearpage

\begin{figure}
\plotone{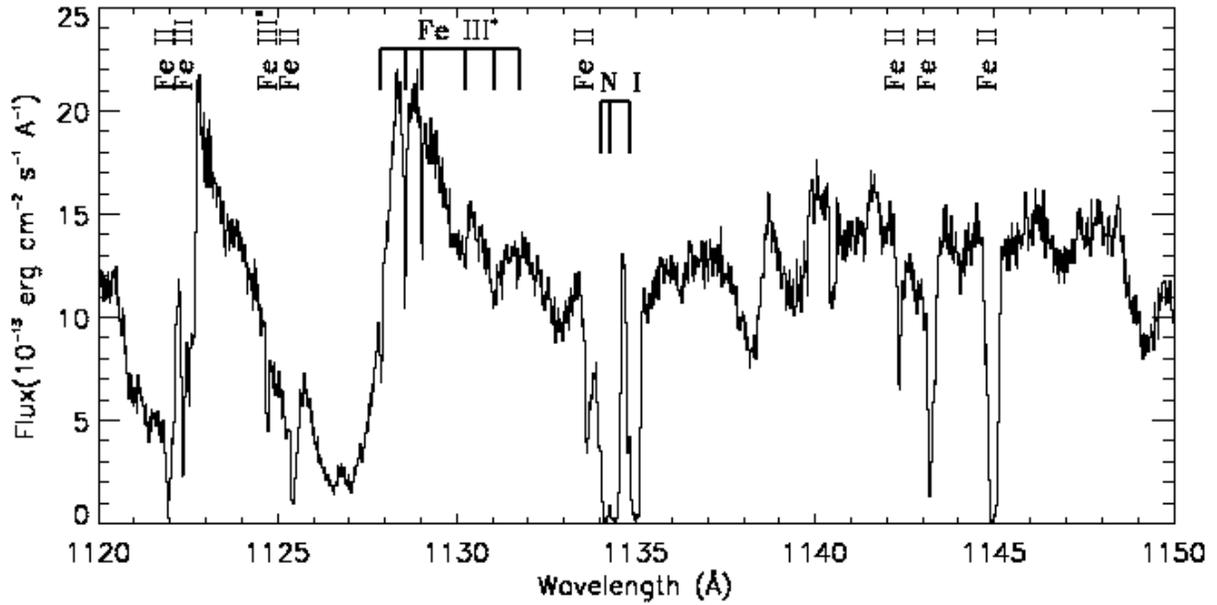}
\caption{A portion of the \emph{FUSE}~LiF~2A spectrum of SwSt~1 from 1120 to 1150 \AA.  The data have been rebinned by three pixels, corresponding to a pixel size of 0.02 \AA.  Nebular absorption lines (which are much narrower than the broad stellar features) are identified above the spectrum; asterisks denote absorption from excited levels.}
\end{figure}

\clearpage

\begin{figure}
\figurenum{2a}
\epsscale{0.7}
\plotone{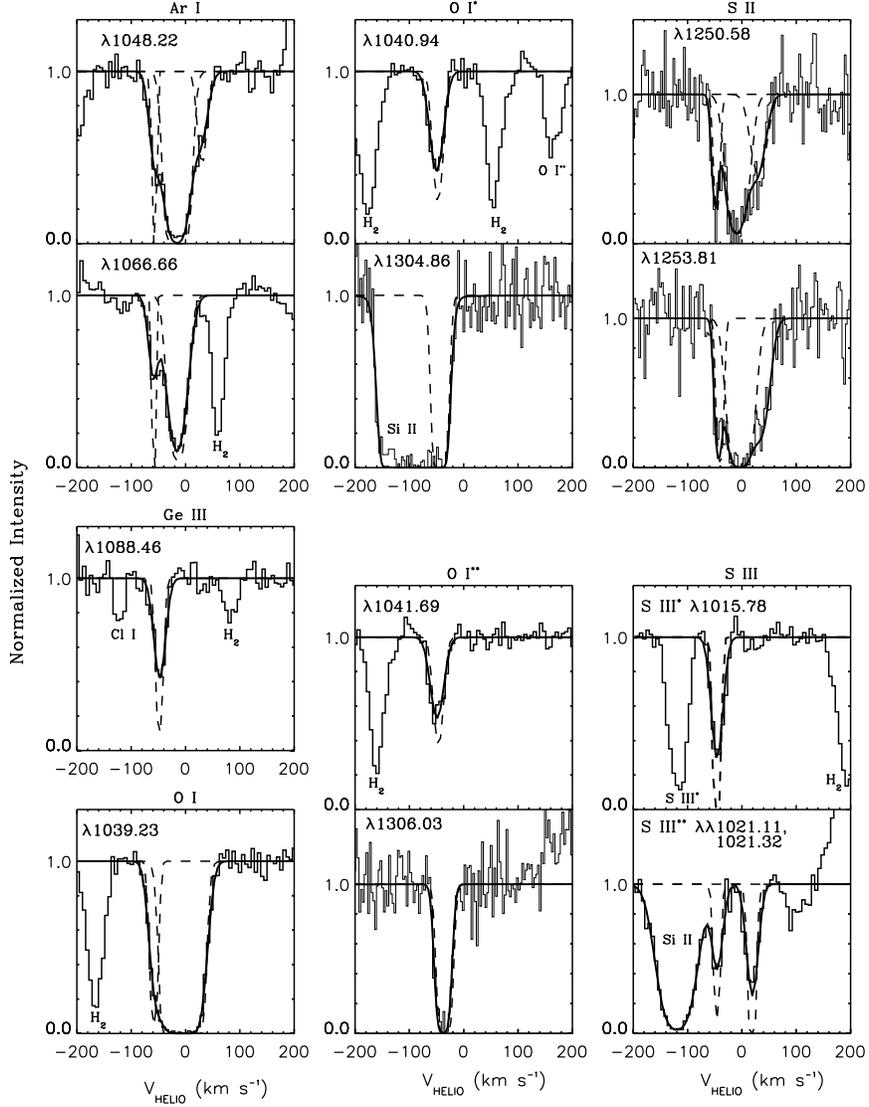}
\caption{Continuum-normalized absorption profiles are plotted versus heliocentric velocity for the identified species.  \ion{S}{2}~$\lambda\lambda$1250.58, 1253.81 and \ion{O}{1}~$\lambda\lambda$1304.86, 1306.03 are from the STIS E140M spectrum of SwSt~1.  All others are from the \emph{FUSE} spectrum: \ion{Ge}{3}~$\lambda$1088.46 is from the LiF~2A segment, while the rest are LiF~1A data.  Double panels are shown for simultaneous fits, and single panels for single line fits.  The measured ion is indicated above the panel (with asterisks signifying absorption from excited levels), and wavelengths are shown within each panel.  Model fits to each profile, convolved with the instrumental spread function, are depicted as thick solid lines.  Thin dashed lines represent fits for the individual components \emph{before} instrumental broadening.  Nebular absorption occurs near $-45$~km~s$^{-1}$.  Other species present in the profiles are indicated.  See text and Table~1 for further information about the displayed fits.}
\end{figure}

\clearpage

\begin{figure}
\figurenum{2b}
\epsscale{0.9}
\plotone{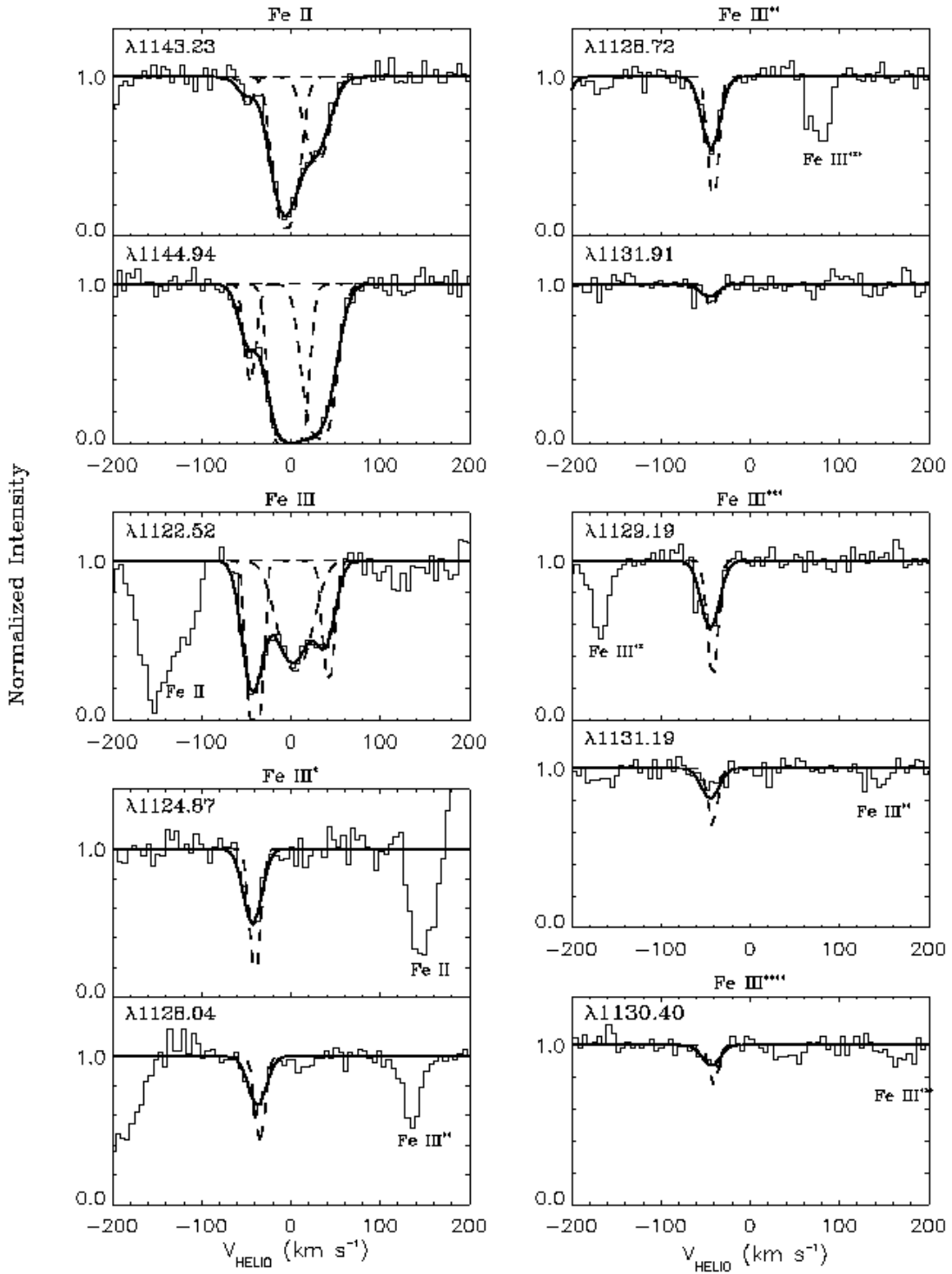}
\caption{Same as Figure~2a, except the displayed continuum-normalized profiles are of iron lines from the LiF~2A segment of the \emph{FUSE} spectrum.}
\end{figure}

\clearpage

\begin{figure}
\figurenum{3}
\epsscale{0.65}
\plotone{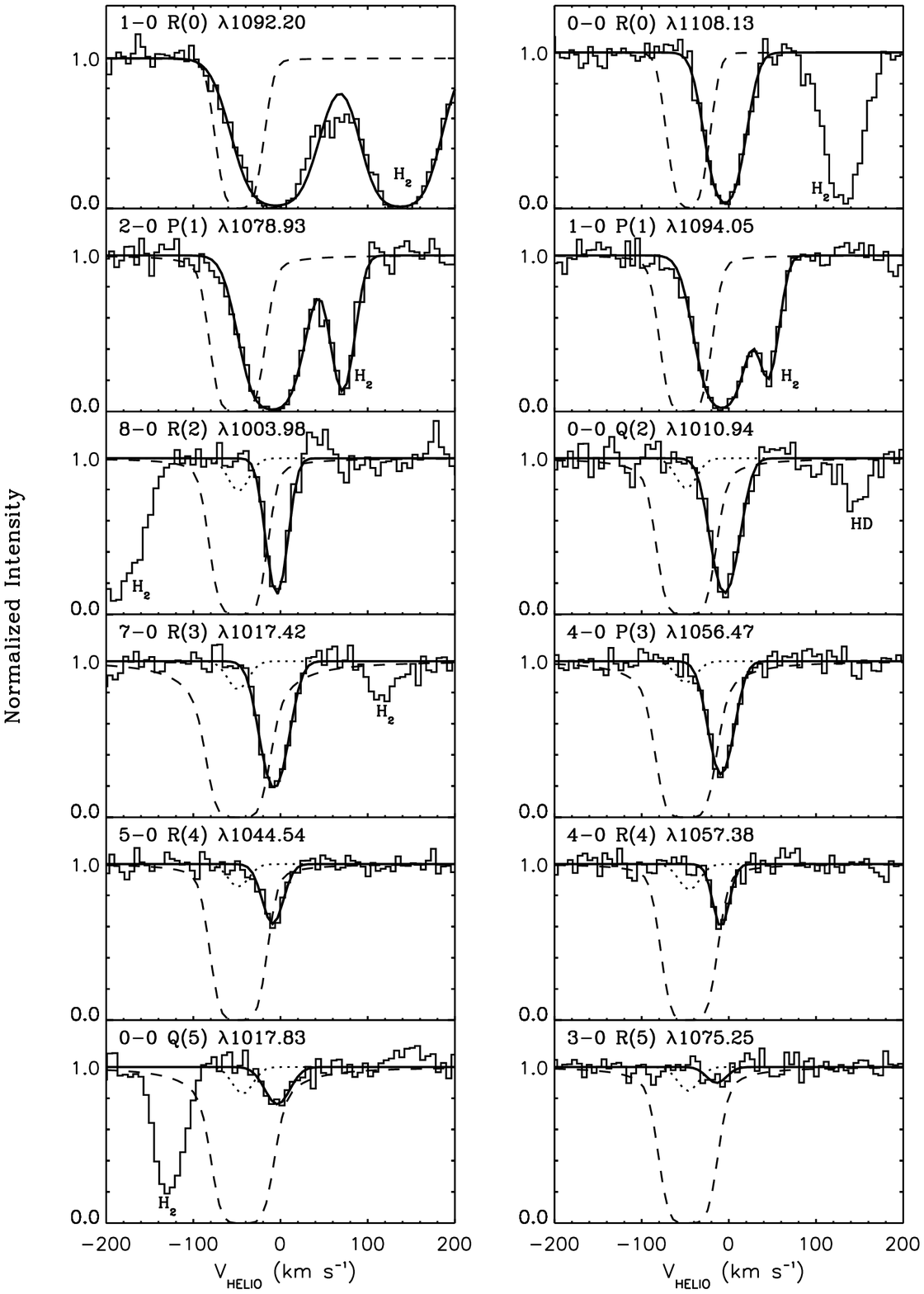}
\caption{Continuum-normalized profiles of selected H$_2$ transitions from Table~7 are plotted against heliocentric velocity.  For comparison, we also display relatively weak (small $f$-value) lines from the $J=0$ and 1 levels, for which no $N_{\rm J}$ upper limits were measured.  Two transitions from each rotational level are shown, and are arranged from top to bottom in order of increasing $J$.  All lines are from the Lyman band, with the exception of the Werner lines 0-0~Q(2)~$\lambda$1010.94 and 0-0~Q(5)~$\lambda$1017.83.  Absorption from other species is indicated within the panels.  Fits to the interstellar component (and blends with nearby lines) are depicted by thick solid lines.  The dotted lines represent nebular absorption profiles with column densities equal to the measured upper limits given in Table~7 (if both $N$(RMS) and $N$(Fit) are available, we select the smallest of the two).  Dashed lines illustrate the absorption profiles expected if the nebular column density of H$_2$ is as high as indicated by the infrared emission lines, assuming $T_{\rm rot}=1600$~K (for lower $T_{\rm rot}$, the nebular absorption profiles would be even more deeply saturated, since the emitting $v=1, J=3$ level would have a smaller fractional population).}
\end{figure}


\begin{thebibliography}{}

\bibitem[Aaquist \& Kwok (1990)]{ak90} Aaquist, O.\ B., \& Kwok, S.\ 1990, A\&AS, 84, 229

\bibitem[Abgrall et al.\ (1993a)]{ab93a} Abgrall, H., Roueff, E., Launay, F., Roncin, J.\ Y., \& Subtil, J.\ L.\ 1993a, A\&AS, 101, 273

\bibitem[Abgrall et al.\ (1993b)]{ab93b} Abgrall, H., Roueff, E., Launay, F., Roncin, J.\ Y., \& Subtil, J.\ L.\ 1993b, A\&AS, 101, 323

\bibitem[Acker et al.\ (2002)]{ack02} Acker, A., Gesicki, K., Grosdidier, Y., \& Durand, S.\ 2002, \aap, 384, 620

\bibitem[Aitken et al.\ (1979)]{ait79} Aitken, D.\ K., Roche, P.\ F., Spenser, P.\ M., \& Jones, B.\ 1979, \apj, 233, 925

\bibitem[Aleman \& Gruenwald (2004)]{alg04} Aleman, I., \& Gruenwald, R.\ 2004, \apj, 607, 865

\bibitem[Aller \& Hyung (1995)]{all95} Aller, L.\ H., \& Hyung, S.\ 1995, \mnras, 276, 1101

\bibitem[Asplund et al.\ (2004)]{asp04} Asplund, M., Grevesse, N., \& Sauval, A.\ J.\ 2004, in Cosmic Abundances as Records of Stellar Evolution and Nucleosynthesis, eds.\ F.\ N.\ Bash \& T.\ G.\ Barnes, ASP Conf.\ Ser., in press (astro-ph/0410214)

\bibitem[Baldwin et al.\ (1996)]{bald96} Baldwin, J.\ A., Crotts, A., Dufour, R.\ J., et~al.\ 1996, \apj, 468, L115

\bibitem[Baluteau et al.\ (1995)]{balu95} Baluteau, J.-P., Zavagno, A., Morisset, C., \& P\'{e}quignot, D.\ 1995, \aap, 303, 175

\bibitem[Bautista \& Kallman (2001)]{bk01} Bautista, M.\ A., \& Kallman, T.\ R.\ 2001, \apjs, 134, 139

\bibitem[Bautista \& Pradhan (1998)]{bp98} Bautista, M.\ A., \& Pradhan, A.\ K.\ 1998, \apj, 492, 650

\bibitem[Bl\"{o}cker (2001)]{bl01} Bl\"{o}cker, T.\ 2001, \apss, 275, 1

\bibitem[Busso et al.\ (1999)]{bus99} Busso, M., Gallino, R., \& Wasserburg, G.\ J.\ 1999, \araa, 37, 239

\bibitem[Busso et al.\ (2001)]{bus01} Busso, M., Gallino, R., Lambert, D.\ L., Travaglio, C., \& Smith, V.\ V.\ 2001, \apj, 557, 802

\bibitem[Casassus et al.\ (2001)]{cas01} Casassus, S., Roche, P.\ F., Aitken, D.\ K., \& Smith, C.\ H.\ 2001, \mnras, 320, 424

\bibitem[Chu et al.\ (2004)]{chu04} Chu, Y.-H., Gruendl, R.\ A., \& Guerrero, M.\ A.\ 2004, in Asymmetrical Planetary Nebulae III, ASP Conf.\ No.\ 313, eds.\ M.\ Meixner et al.\ (San Francisco: ASP), 254

\bibitem[Cohen et al.\ (2002)]{coh02} Cohen, M., Barlow, M.\ J., Liu, X.-W., \& Jones, A.\ F.\ 2002, \mnras, 332, 879

\bibitem[Crowther et al.\ (1998)]{cro98} Crowther, P.\ A., De~Marco, O., \& Barlow, M.\ J.\ 1998, \mnras, 296, 367

\bibitem[de Freitas Pacheco \& Veliz (1987)]{dfp87} de~Freitas~Pacheco, J.\ A., \& Veliz, J.\ G.\ 1987, \mnras, 227, 773 (dFP87)

\bibitem[De Marco et al.\ (2001)]{dm01} De Marco, O., Crowther, P.\ A., Barlow, M.\ J., Clayton, G.\ C., \& de Koter, A.\ 2001, \mnras, 328, 527 (DM01)

\bibitem[De Marco \& Soker(2002)]{ds02} De Marco, O., \& Soker, N.\ 2002, \pasp, 114, 602

\bibitem[Dinerstein et al.\ (1988)]{din88} Dinerstein, H.\ L., Lester, D.\ F., Carr, J.\ S., \& Harvey, P.\ M.\ 1988, \apj, 327, L27

\bibitem[Dinerstein \& Sneden(1988)]{ds88} Dinerstein, H.\ L.\ \& Sneden, C.\ 1988, \apj, 335, L23

\bibitem[Dinerstein et al.\ (1995a)]{din95a} Dinerstein, H.\ L., Haas, M.\ R., Erickson, E.\ F., \& Werner, M.\ W.\ 1995a, in Airborne Astronomy Symposium on the Galactic Ecosystem, ASP Conf.\ Ser.\ No.\ 73, eds.\ M.\ R.\ Haas, J.\ A.\ Davidson, \& E.\ F.\ Erickson (San Francisco: ASP), 387

\bibitem[Dinerstein et al.\ (1995b)]{din95b} Dinerstein, H.\ L., Sneden, C., \& Uglum, J.\ 1995b, \apj, 447, 262 (DSU)

\bibitem[Dinerstein et al.\ (1997)]{din97} Dinerstein, H.\ L., Haas, M.\ R., Erickson, E.\ F., \& Werner, M.\ W.\ 1997, in IAU Symposium 180: Planetary Nebulae, eds.\ H.\ J.\ Habing \& H.\ J.\ G.\ L.\ M.\ Lamers (Dordrecht: Kluwer), 220

\bibitem[Dinerstein \& Bowers(2004)]{db04} Dinerstein, H.\ L.\ \& Bowers, C.\ W.\ 2004, in Asymmetrical Planetary Nebulae III, eds.\ M.\ Meixner et al.\ ASP Conf.\ No.\ 313 (San Francisco: ASP), 347

\bibitem[Dorfi \& H\"{o}fner (1996)]{dh96} Dorfi, E.\ A., \& H\"{o}fner, S.\ 1996, \aap, 313, 605

\bibitem[Draine (2003)]{dr03} Draine, B.\ T.\ 2003, \araa, 41, 241

\bibitem[Dwarkadas \& Balick (1998)]{dwb98} Dwarkadas, V.\ V., \& Balick, B.\ 1998, \apj, 497, 267

\bibitem[Dyson et al.\ (1989)]{dy89} Dyson, J.\ E., Hartquist, T.\ W., Pettini, M., \& Smith, L.\ J.\ 1989, \mnras, 241, 625

\bibitem[Ekberg (1993)]{ek93} Ekberg, J.\ O.\ 1993, A\&AS, 101, 1

\bibitem[Esteban et al.\ (2004)]{est04} Esteban, C., Peimbert, M., Garc\'{i}a-Rojas, J., Ruiz, M.\ T., Peimbert, A., \& Rodr\'{i}guez, M.\ 2004, \mnras, 355, 229

\bibitem[Flower et al.\ (1984)]{flo84} Flower, D.\ R., Goharji, A., \& Cohen, M.\ 1984, \mnras, 206, 293

\bibitem[Frank (1995)]{fr95} Frank, A.\ 1995, \aj, 110, 2457

\bibitem[Galavis et al.\ (1995)]{gal95} Galavis, M.\ E., Mendoza, C., \& Zeippen, C.\ J.\ 1995, A\&AS, 111, 347

\bibitem[Garc\'{i}a-Segura \& Franco (1996)]{gsf96} Garc\'{i}a-Segura, G., \& Franco, J.\ 1996, \apj, 469, 171

\bibitem[Garstang et al.\ (1978)]{gar78} Garstang, R.\ H., Robb, W.\ D., \& Rountree, S.\ P.\ 1978, \apj, 222, 384

\bibitem[Gon\c{c}alves et al.\ (2001)]{gon01} Gon\c{c}alves, D.\ R., Corradi, R.\ L.\ M., \& Mampaso, A.\ 2001, \apj, 547, 302

\bibitem[Goriely \& Mowlavi (2000)]{gor00} Goriely, S., \& Mowlavi, N.\ 2000, \aap, 362, 599

\bibitem[G\'{o}rny \& Tylenda (2000)]{gt00} G\'{o}rny, S.\ K., \& Tylenda, R.\ 2000, \aap, 362, 1008

\bibitem[Grandi (1975a)]{gr75a} Grandi, S.\ A.\ 1975a, \apj, 196, 465

\bibitem[Grandi (1975b)]{gr75b} Grandi, S.\ A.\ 1975b, \apj, 199, L43

\bibitem[Grandi (1976)]{gr76} Grandi, S.\ A.\ 1976, \apj, 206, 658

\bibitem[Grosdidier et al.\ (2000)]{gros00} Grosdidier, Y., Acker, A., \& Moffat, A.\ F.\ J.\ 2000, \aap, 364, 597

\bibitem[Guhathakurta \& Draine (1989)]{gud89} Guhathakurta, P., \& Draine, B.\ T.\ 1989, \apj, 345, 230

\bibitem[Gussie \& Taylor(1995)]{gus95} Gussie, G.\ T.\ \& Taylor, A.\ R.\ 1995, \mnras, 273, 801

\bibitem[Herwig et al.\ (1999)]{her99} Herwig, F., Bl\"{o}cker, T., Langer, N., \& Drieve, T.\ 1999, \aap, 349, L5

\bibitem[Herwig et al.\ (2003)]{her03} Herwig, F., Langer, N., \& Lugaro, M.\ 2003, \apj, 593, 1056

\bibitem[Hoare et al.\ (1992)]{hrc92} Hoare, M.\ G., Roche, P.\ F., \& Clegg, R.\ E.\ S.\ 1992, \mnras, 258, 257

\bibitem[Hollenbach \& Tielens(1997)]{hol97} Hollenbach, D.\ J., \& Tielens, A.\ G.\ G.\ M.\ 1997, \araa, 35, 179

\bibitem[Huggins et al.(1996)]{hug96} Huggins, P.\ J., Bachiller, R., Cox, P., \& Forveille, T.\ 1996, \aap, 315, 284

\bibitem[Huggins et al.\ (2002)]{hug02} Huggins, P.\ J., Forveille, T., Bachiller, R., Cox, P., Ageorges, N., \& Walsh, J.\ 2002, \apj, 573, L55

\bibitem[Huggins \& Mauron (2002)]{hm02} Huggins, P.\ J., \& Mauron, N.\ 2002, \aap, 393, 273

\bibitem[Hummer et al.\ (1993)]{hum93} Hummer, D.\ G., Berrington, K.\ A., Eissner, W., Pradhan, A.\ K., Saraph, H.\ E., \& Tully, J.\ A.\ 1993, \aap, 279, 298

\bibitem[Hyung et al.\ (1994)]{hyu94} Hyung, S., Aller, L.\ H., \& Feibelman, W.\ A.\ 1994, \mnras, 269, 975

\bibitem[Kallman \& Bautista (2001)]{kb01} Kallman, T., \& Bautista, M.\ 2001, \apjs, 133, 221

\bibitem[Keenan et al.\ (1992)]{kee92} Keenan, F.\ P., Berrington, K.\ A., Burke, P.\ G., Zeippen, C.\ J., Le~Dourneuf, M., \& Clegg, R.\ E.\ S.\ 1992, \apj, 384, 385

\bibitem[Keenan et al.\ (1993)]{kee93} Keenan, F.\ P., Aller, L.\ H., Hyung, S., Conlon, E.\ S., \& Warren, G.\ A.\ 1993, \apj, 410, 430

\bibitem[Keenan et al.\ (2001)]{kee01} Keenan, F.\ P., Aller, L.\ H., Ryans, R.\ S.\ I., \& Hyung, S.\ 2001, Proc.\ Nat.\ Acad.\ Sci., 98, 9476

\bibitem[Kimble et al.\ (1998)]{kim98} Kimble, R.\ A.\ et~al.\ 1998, \apj, 492, L83

\bibitem[Kingdon et al.\ (1995)]{kff95} Kingdon, J., Ferland, G.\ J., \& Feibelman, W.\ A.\ 1995, \apj, 439, 793

\bibitem[Kingdon \&  Ferland (1997)]{kf97} Kingdon, J.\ B., \& Ferland, G.\ J., 1997, \apj, 477, 732

\bibitem[Kwok et al.\ (1981)]{kw81} Kwok, S., Purton, C.\ R., \& Keenan, D.\ W.\ 1981, \apj, 250, 232

\bibitem[Latter et al.\ (1995)]{lat95} Latter, W.\ B., Kelly, D.\ M., Hora, J.\ L., \& Deutsch, L.\ K.\ 1995, \apjs, 100, 159

\bibitem[Likkel et~al.\ 2004]{lik04} Likkel, L., Bruch, J., Bartig, K., Dinerstein, H.\ L., \& Lester, D.\ F.\ 2004, in Asymmetric Planetary Nebulae III, ed.\ M.\ Meixner et~al., ASP Conf.\ No.\ 313, 351

\bibitem[Lindler (1999)]{lind99} Lindler, D.\ 1999, CALSTIS Reference Guide (Greenbelt: NASA/LASP)

\bibitem[Liseau et al.\ (1999a)]{lis99a} Liseau, R., et al.\ 1999a, \aap, 344, 342

\bibitem[Liseau et al.\ (1999b)]{lis99b} Liseau, R., White, G.\ J., \& Larsson, B.\ 1999b, in The Universe as Seen by ISO, ESA SP-427, eds.\ P.\ Cox \& M.\ F.\ Kessler, 703

\bibitem[Liu et al.\ (2001)]{liu01} Liu, X.-W., et al.\ 2001, \mnras, 323, 343

\bibitem[Lucy (1995)]{lucy} Lucy, L.\ 1995, \aap, 294, 555

\bibitem[McCandliss (2003)]{mcc03} McCandliss, S.\ R.\ 2003, \pasp, 115, 651

\bibitem[Mendoza \& Zeippen (1982)]{men82} Mendoza, C., \& Zeippen, C.\ J.\ 1982, \mnras, 199, 1025

\bibitem[Middlemass (1988)]{mid88} Middlemass, D.\ 1988, \mnras, 231, 1025

\bibitem[Mizutani et al.\ (2004)]{miz04} Mizutani, M., Onaka, T., \& Shibai, H.\ 2004, \aap, 423, 579

\bibitem[Moos et al.\ (2000)]{moo00} Moos, H.\ W., et~al.\ 2000, \apj, 538, L1

\bibitem[Morton (1991)]{mor91} Morton, D.\ C.\ 1991, \apjs, 77, 119

\bibitem[Morton (2000)]{mor00} Morton, D.\ C.\ 2000, \apjs, 130, 403

\bibitem[Morton (2003)]{mor03} Morton, D.\ C.\ 2003, \apjs, 149, 205

\bibitem[Nahar (1993)]{nah93} Nahar, S.\ N.\ 1993, \physscr, 48, 297

\bibitem[Nahar \& Pradhan (1996)]{np96} Nahar, S.\ N., \& Pradhan, A.\ K.\ 1996, A\&AS, 119, 509 (NP96)

\bibitem[O'Dell et al.\ (2002)]{ode02} O'Dell, C.\ R., Balick, B., Haijan, A.\ R., Henney, W.\ J., \& Burkert, A.\ 2002, \aj, 123, 3329

\bibitem[Pe\~{n}a et al.\ (2003)]{pen03} Pe\~{n}a, M., Medina, S., \& Stasi\'{n}ska, G.\ 2003, RMxA\&A (Ser.\ de Conf.), 18, 84

\bibitem[Perinotto et al.\ (1999)]{per02} Perinotto, M., Bencini, C.\ G., Pasquali, A., Manchado, A., Rodriguez Espinosa, J.\ M., \& Stanga, R.\ 1999, \aap, 347, 967

\bibitem[Ramsay et al.\ (1993)]{ram93} Ramsay, S.\ K., Chrysostomou, A., Geballe, T.\ R., Brand, P.\ W.\ J.\ L., \& Mountain, M.\ 1993, \mnras, 263, 695

\bibitem[Roche et al.\ (1996)]{roc96} Roche, P.\ F., Lucas, P.\ W., Hoare, M.\ G., Aitken, D.\ K., \& Smith, C.\ H.\ 1996, \mnras, 280, 924

\bibitem[Rodr\'{i}guez (1999)]{rod99} Rodr\'{i}guez, M.\ 1999, \aap, 348, 222

\bibitem[Rodr\'{i}guez (2002)]{rod02} Rodr\'{i}guez, M.\ 2002, \aap, 389, 556

\bibitem[Rudy et al.\ (1989)]{rud99} Rudy, R.\ J., Rossano, G.\ S., \& Puetter, R.\ C.\ 1989, \apj, 346, 799

\bibitem[Rudy et al.\ (1991a)]{rud91a} Rudy, R.\ J., Rossano, G.\ S., Erwin, P., \& Puetter, R.\ C.\ 1991a, \apj, 368, 468

\bibitem[Rudy et al.\ (1991b)]{rud91b} Rudy, R.\ J., Cohen, R.\ D., Rossano, G.\ S., Erwin, P., Puetter, R.\ C., \& Lynch, D.\ K.\ 1991b, \apj, 380, 151

\bibitem[Rudy et al.\ (1992)]{rud92} Rudy, R.\ J., Erwin, P., Rossano, G.\ S., \& Puetter, R.\ C.\ 1992, \apj, 384, 536

\bibitem[Sahnow et al.\ (2000)]{sah00} Sahnow, D.\ J., et~al.\ 2000, \apj, 538, L7

\bibitem[Savage et al.\ (1977)]{sav77} Savage, B.\ D., Drake, J.\ F., Budich, W., \& Bohlin, R.\ C.\ 1977, \apj, 216, 291

\bibitem[Savage \& Sembach (1996)]{sav96} Savage, B.\ D., \& Sembach, K.\ R.\ 1996, \araa, 34, 279

\bibitem[Sembach \& Savage (1996)]{ss96} Sembach, K.\ R.\, \& Savage, B.\ D.\ 1996, \apj, 457, 211

\bibitem[Sharpee et al.\ (2003)]{sharp03} Sharpee, B., Williams, R., Baldwin, J.\ A., \& van~Hoof, P.\ A.\ M.\ 2003, \apjs, 149, 157

\bibitem[Shaw \& Dufour (1995)]{sh95} Shaw, R.\ A., \& Dufour, R.\ J.\ 1995, \pasp, 107, 896

\bibitem[Shields (1978)]{shi78} Shields, G.\ A.\ 1978, \apj, 219, 559

\bibitem[Shull et al.\(2004)]{shl04} Shull, J.\ M., et al.\ 2004, \baas, 204, \#61.18

\bibitem[Siess et al.\ (2004)]{sie04} Siess, L., Goriely, S., \& Langer, N.\ 2004, \aap, 415, 1089

\bibitem[Soker \& Clayton (1999)]{sok99} Soker, N., \& Clayton, G.\ C.\ 1999, \mnras, 307, 993

\bibitem[Speck et al.\ (2003)]{spe03} Speck, A.\ K., Meixner, M., Jacoby, G.\ H., \& Knezek, P.\ 2003, \pasp, 115, 170

\bibitem[Sterling et al.\ (2002)]{ster02} Sterling, N.\ C., Dinerstein, H.\ L., \& Bowers, C.\ W.\ 2002, \apj, 578, L55 (SDB)

\bibitem[Sterling \& Dinerstein (2003)]{ster03} Sterling, N.\ C., \& Dinerstein, H.\ L.\ 2003, RMxA\&A (Ser.\ de Conf.), 18, 133

\bibitem[Swings \& Struve (1940)]{swst40} Swings, P., \& Struve, O.\ 1940, Proc.\ Nat.\ Acad.\ Sci., 26, 454

\bibitem[Szczerba et al.\ (2001)]{sz01} Szczerba, R., G\'{o}rny, S.\ K., Stasi\'{n}ska, G., Si\'{o}dmiak, N., \& Tylenda, R.\ 2001, \apss, 275, 113

\bibitem[Tielens \& Hollenbach (1985)]{tie85} Tielens, A.\ G.\ G.\ M., \& Hollenbach, D.\ 1985, \apj, 291, 722

\bibitem[Tielens (1993)]{tie93} Tielens, A.\ G.\ G.\ M.\ 1993, in Planetary Nebulae: IAU Symposium 155, eds.\ R.\ Weinberger \& A.\ Acker (Dordrecht: Kluwer), 155

\bibitem[Tylenda et al.\ (1993)]{tyl93} Tylenda, R., Acker, A., \& Stenholm, B.\ 1993, \aaps, 102, 595

\bibitem[Watson(1984)]{wat84} Watson, D.\ M.\ 1984, in Galactic and  Extragalactic Infrared Spectroscopy, eds.\ M.\ F.\ Kessler \& J.\ P.\ Phillips (Dordrecht: Reidel), 195

\bibitem[Welty et al.\ (1999)]{wel99} Welty, D.\ E., Hobbs, L.\ M., Lauroesch, J.\ T., Morton, D.\ C., Spitzer, L., \& York, D.\ G.\ 1999, \apjs, 124, 465

\bibitem[Woodgate et al.\ (1998)]{wood98} Woodgate, B.\ E., et~al.\ 1998, \pasp, 110, 1183

\bibitem[Zhang \& Kwok (1990)]{zk90} Zhang, C.\ Y., \& Kwok, S.\ 1990, \aap, 237, 479

\bibitem[Zhang (1996)]{zh96} Zhang, H.\ L.\ 1996, A\&AS, 119, 523

\bibitem[Zhang \& Liu (2002)]{zh02} Zhang, Y., \& Liu, X.-W.\ 2002, \mnras, 337, 499

\end{thebibliography}
\end{document}